\documentclass[10pt,journal,compsoc]{IEEEtran}
\ifCLASSOPTIONcompsoc
  \usepackage[nocompress]{cite}
\else
  \usepackage{cite}
\fi
\ifCLASSINFOpdf
\else
\fi
\usepackage{booktabs} 
\usepackage{multirow}
\usepackage{graphicx}
\usepackage{enumerate}
\usepackage{subfigure}
\usepackage{threeparttable}
\usepackage{enumerate}
\usepackage{amssymb}
\usepackage[ruled,linesnumbered]{algorithm2e}
\usepackage{amsmath}
\usepackage{float}
\usepackage{setspace}
\usepackage{enumitem}
\usepackage{color,xcolor}
\usepackage{verbatim}
\usepackage{balance}
\usepackage{soul}
\usepackage{url}
\usepackage{hyperref}
\begin{document}
%
\title{Incorporating Interactive Facts for Stock Selection via Neural Recursive ODEs}
%
%
%
%

\author{
        Qiang~Gao,
        Xinzhu~Zhou,
        Kunpeng~Zhang,
        Li~Huang,
        Siyuan Liu,
        and Fan~Zhou
\thanks{Li~Huang is the corresponding author.}
\IEEEcompsocitemizethanks{
\IEEEcompsocthanksitem Q.~Gao, X.~Zhou, and L.~Huang are with the School of Computing and Artificial Intelligence, Southwestern University of Finance and Economics, Chengdu, China, 611130. \protect 
\\ E-mail: qianggao@swufe.edu.cn, 222020204197@smail.swufe.edu.cn,\\ lihuang@swufe.edu.cn
\IEEEcompsocthanksitem K. Zhang is with the University of Maryland, College Park, USA. 
\\E-mail: kpzhang@umd.edu
\IEEEcompsocthanksitem S. Liu is with the Pennsylvania State University, PA 16802 USA. 
\\E-mail: 
siyuan@psu.edu.
\IEEEcompsocthanksitem F. Zhou is with the School of Information and Software Engineering, University of Electronic Science and Technology of China, China.
\\E-mail: fan.zhou@uestc.edu.cn
}
\thanks{Manuscript received XX XX, 2022; revised XX XX, 2022.}\\
}

%
%

\markboth{IEEE Transactions on XXX,~Vol.~XX, No.~XX, November~2022}%
{Shell \MakeLowercase{\textit{Gao et al.}}: Incorporating Interactive Facts for Stock Selection via Neural Recursive ODEs}
%



\IEEEtitleabstractindextext{%
\begin{abstract}
Stock selection attempts to rank a list of stocks for optimizing investment decision making, aiming at minimizing investment risks while maximizing profit returns. Recently, researchers have developed various (recurrent) neural network-based methods to tackle this problem. Without exceptions, they primarily leverage historical market volatility to enhance the selection performance. However, these approaches greatly rely on discrete sampled market observations, which either fail to consider the uncertainty of stock fluctuations or predict continuous stock dynamics in the future. Besides, some studies have considered the explicit stock interdependence derived from multiple domains (e.g., industry and shareholder). Nevertheless, the implicit cross-dependencies among different domains are under-explored. To address such limitations, we present a novel stock 
selection solution -- StockODE, a latent variable model with Gaussian prior. Specifically, we devise a Movement Trend Correlation module to expose the time-varying relationships regarding stock movements. We design Neural Recursive Ordinary Differential Equation Networks (NRODEs) to capture the temporal evolution of stock volatility in a continuous dynamic manner. Moreover, we build a hierarchical hypergraph to incorporate the 
domain-aware dependencies among the stocks. Experiments conducted on two real-world stock market datasets demonstrate that StockODE significantly outperforms several baselines, such as up to 18.57\% average improvement regarding Sharpe Ratio.
\end{abstract}

}

\maketitle

\IEEEdisplaynontitleabstractindextext

%
\IEEEpeerreviewmaketitle

\section{Introduction}
\label{Introduction}
The continual growth of global market capitalization has spawned the prosperity of quantitative investment, which provides numerous investors and traders unprecedented opportunities to meticulously select valuable stocks for maximizing profit returns. The majority of ground-breaking research focuses on stock prediction tasks that can be formulated as binary classification or regression tasks to forecast future stock movements (e.g. stock prices)~\cite{chen2015lstm, Feng2019TemporalRR}. In contrast, this study investigates a newly emerging but more complex task, i.e., stock selection, which can be framed as a \textit{learning to rank} problem. Specifically, stock selection mainly attempts to optimize the target of investment regarding the profit returns, which not only considers the future trends of massive stock candidates but also needs to select the stocks with maximum profitable returns. Consequently, tackling the stock selection problem is a challenging and non-trivial task owing to several uncertainties such as high volatility and historical dependence on the stock market.

As one of the common clues, learning historical stock transactions can help reveal the stock's future evolution, which in turn enables us to determine which stock is the most profitable investment. The motivation is that earlier studies have demonstrated that stock movements, to some extent, are predictable according to literature in behavioral finance and behavioral economics~\cite{nofsinger2005social,cheng2021modeling}. Conventional approaches usually adopt time-series models to capture the historical influences and changes of diverse time-series market signals, such as open price, close price, trading volume, and others. For instance, researchers have used ARIMA~\cite{ariyo2014stock} or GARCH~\cite{franses1996forecasting} to study the stock volatility based on a given indicator, e.g., daily close price. Moreover, numerous endeavors develop various machine learning algorithms, e.g., SVM~\cite{lin2013svm}, Random Forest~\cite{Sadorsky2021ARF}, HMM~\cite{hassan2005stock}, to explore the stock fluctuations from historical transaction data. However, those approaches fail to take into account the inherent non-stationary fluctuations of the stock market and are not usually compatible with stock trends in real financial scenarios, which could ultimately fail in choosing more profitable stocks~\cite{wang2021}. 

Recently, the widely used deep learning approaches such as recurrent neural networks have become a natural choice for capturing the evolution of historical stock transactions~\cite{chen2015lstm,zhang2017stock}. In addition, The Efficient Market Hypothesis (EMH) indicates that the stock movements could be affected by relevant information from multiple sources~\cite{fama1965behavior}. For instance, textual data sources such as financial news and public reviews are now being used to expose their influence in understanding the stock 
trends~\cite{shi2018deepclue,li2020multimodal,sawhney2021fast}. Publicly available company information, as another source has been analyzed to investigate the internal dependencies in the stock market. It has become a substantial and valuable clue to promote stock movement modeling in addition to solely exploring the historical stock moving patterns~\cite{gao2021graph,cheng2021modeling}. Recently, deep graph learning-based studies have demonstrated that modeling the historical stock movements along with taking into account the stock dependencies can significantly help predict near-term stock trends, which further 
enables us to select or recommend profitable stocks for investment~\cite{Feng2019TemporalRR,wang2021}. We position our study in the second direction, as it focuses more on the implied and interactive information behind the stocks. 
\begin{figure}[t]
    \centering
    \includegraphics[width=0.5\textwidth]{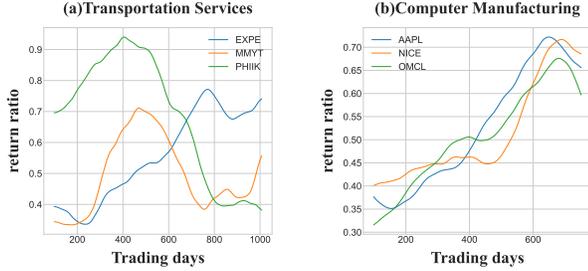}
    \caption{An example of stock's movement regarding return ratio. The changes in return ratios for various stock market tickers are shown by the easily distinct colored curves.}
    \label{fig:stockmove}
\end{figure}

Although graph-based techniques have received a great deal of attention for understanding stock relationships, we consider that there are still two main interactive facts between stocks that may affect the capture of potential interdependencies: \textit{(a) Time-varying Correlation.} As shown in Fig.~\ref{fig:stockmove}, we respectively visualize the trend of one-day returns for three stocks in two different industries. We can find that the correlation strengths between these stocks evolve over time. Some stocks (e.g., \textit{APPL} and \textit{NICE}) have synchronous evolving trends while some (e.g., \textit{EXPE} and \textit{PHIIK}) maintain competitive trends. In the real world, this phenomenon is prevalent in stock markets due to the Momentum Spillover Effect~\cite{cheng2021modeling}. For instance, the stock price of Intel typically has an impact on the stock prices of other semiconductor firms. However, existing studies usually treat each stock's dynamics as isolated, adhering to the standard (deep learning-based) time series modeling paradigm~\cite{shen2018deep,ding2020hierarchical}, and do not consider the interactive signals between stocks regarding their past movements. This makes it impossible to specify how stock correlations have changed over time. Therefore, We consider that incorporating such time-varying correlations behind the stock movements could boost the understanding of distinct stock volatility trends. 
\textit{(b) Domain-aware Dependency.} Despite the fact that predetermined stock relationships can be extracted from numerous publicly available domains (e.g., industry and shareholder) to enhance the stock selection performance owing to the powerful capability of GNN-based models, such as~\cite{ying2020time,sawhney2021exploring}, we argue that existing solutions only exploit the simple stock relations within a specific domain (e.g., the stocks from the same industry) while the underlying cross-dependencies among different domains are under-explored. For instance, \textit{United Airlines}, \textit{FedEx} and \textit{Delta Air Lines} all operate \textit{Boeing} planes, but \textit{Delta Air Lines} has a stronger relationship with \textit{United Airlines} than \textit{FedEx} and \textit{Delta Air Lines} as \textit{Delta Air Lines} and \textit{United Airlines} belong to the same industry. Thus, we consider that employing shallow knowledge can only bring limited benefits to selection performance as the complex higher-order collaborations on the predefined stock relations are not fully explored.

In addition, stock selection, as a typical time-dependent learning task, naturally includes a series of indicator observations that reflect stock trends. Furthermore, stock movements may change significantly in a short period, e.g., significant volatility in the hours after the stock market opens~\cite{chen2018neural,lu2018beyond}. Existing approaches based on deep neural networks should have the capability of modeling time-series data in continuous dynamic systems to address stock volatility. However, in practice, they can only take the discrete-time observations (e.g., daily level) as the input and use the static topological information structure to predict the most profitable stocks, which could result in the gap between deep neural networks and dynamic systems, ignore the uncertainty behind stock observations, and even cause the failure of predicting continuous-time dynamics regarding stock movements. For example, we usually employ RNN-based models to handle the historical price data derived from daily observations, whereafter we use them to predict future stock trends at a fixed time interval~\cite{shen2018deep,qin2017dual}. However, these models could lead to investment risk and financial loss when the stock prices fluctuate rapidly within minutes or hours~\cite{scholtus2014speed,sawhney2021fast}.

To remedy the aforementioned drawbacks, we introduce a novel framework named \textbf{StockODE} to solve the stock selection problem, which is motivated by recent successful applications of Neural Ordinary Differential Equations (NODEs) in time-series data~\cite{chen2018neural,zhou2021forecasting}. In detail, we first develop a Movement Trend Correlation module to initiate the capture of time-varying correlations between different stocks' historical movements (e.g., return ratio). Next, we design an attention-inspired Neural Recursive ODE (NRODE) block to model the historical multivariate stock movements in a continuous dynamic manner. In particular, StockODE performs as a latent variable model to relieve the uncertainty of stock fluctuations via Gaussian assumption. Finally, a Hierarchical Hypergraph Convolution Network (HHCN) is presented to address the domain-aware dependencies among the stocks extracted from relational knowledge of market environmental variables. Notably, our hierarchical hypergraph network constructs the stock interactions from intra-domain knowledge and cross-domain knowledge to learn the complexly higher-order interactions among the stocks. The main contributions of our study are summarized as follows:
\begin{itemize}
    \item We propose a flexibly dynamic neural framework--StockODE, which provides a new perspective on stock movement learning. To the best of our knowledge, StockODE is the first attempt to involve the neural ODE to capture the temporal evolution of stock volatility in a continuous dynamic manner.
    \item We design a Movement Trend Correlation module that is capable of integrating the co-evolution and anti-evolution relationships between different stocks regarding historical movements.
    \item To capture the domain-aware dependencies behind various stocks, we devise a hierarchical hypergraph to describe the intra-domain and inter-domain knowledge from real-world relation sources.
    \item Our experimental results conducted on two real-world datasets demonstrate that the proposed StockODE significantly outperforms the existing solutions regarding quantitative stock trading.
\end{itemize}

In the rest of this paper,
we first review the relevant studies 
in Section~\ref{Related Work}, and 
then formalize the problem in Section~\ref{Problem Definition}. The details of the proposed \textbf{StockODE} framework are discussed in Section~\ref{Methodology}, and 
the results of the experimental evaluations quantifying the benefits of our approach are presented in Section~\ref{Experiments}. Section~\ref{sec:concl} concludes this study and  
outlines the directions of future work.

\section{Related Work}
\label{Related Work}
We divide the relevant research into three key categories, including conventional stock movement learning, deep learning-based approaches, and market relation learning.

\subsection{Conventional Stock Movement Learning}
The early efforts attempt to leverage technical analysis and fundamental analysis for stock market modeling, where the former mainly relies upon the past stock trends as future indicators while the latter aims to investigate the intrinsic value of stock price, i.e., fair value~\cite{jiang2021applications}. For technical analysis, it aims at extracting the volume indicators from historical stock movement data, whereafter adopting the linear models, e.g., ARIMA~\cite{ariyo2014stock}, to predict stock price trends. Moreover, several machine learning-based methods, e.g., SVM and Random Forest, have been successfully involved in learning stock dynamics~\cite{lin2013svm,Sadorsky2021ARF}. In contrast to technical analysis, the fundamental analysis presents a newer perspective for the stock movement prediction, which considers the impact factors from the third-party data, such as social media, earning calls, and financial news~\cite{nguyen2015topic,wang2020incorporating}. 
In essence, the existing studies concentrate on the stock trend prediction task and use a classification or regression scheme to infer the binary trend or stock future price. For instance, researchers exploit the volume indicators of historical transaction data with respect to stock movement and endeavor to model the stock dynamics with popular machine learning approaches such as Logistic Regression~\cite{dutta2012prediction}. In contrast, stock selection (or ranking/recommendation) aims at providing optimal stock choices by ranking a stock list, which helps in achieving more profit expectations as well as relieving the investment risks~\cite{sawhney2021stock,gao2021graph}.

\subsection{Deep Learning-based Approaches}
Since the stock market is a dynamic system affected by multiple time-varying signals, the emerging deep neural networks especially for the recurrent networks, e.g., LSTM (long-short term memory)~\cite{Hochreiter1997} and GRU (gated recurrent units)~\cite{Chung2014}, have the capability of capturing the intricate temporal dependencies behind the multivariate time series transactions, which spurs the researchers to apply them for historical stock signal modeling~\cite{cheng2018applied}. For instance, \cite{chen2015lstm} extended the LSTM model to enhance the accuracy of stock return prediction. \cite{nelson2017stock} predicted future trends of stock prices based on the price history and technical analysis indicators, whereafter it concluded that the RNN-based models achieve significant gains compared to the earlier machine learning approaches. Other popular methods, e.g., seq2seq-based~\cite{wang2021clvsa,xu2018stock}, attention-based~\cite{wang2021,ding2020hierarchical,wang2022hatr}, have also attracted researchers' interest in stock movement learning. Due to the inherent limitation of existing deep recursive neural networks that can only receive or predict discrete time-series signals, we argue that modeling stock market should have the capability of adapting to the continuous dynamic scenarios, enhancing the perception of market volatility or uncertainty. Motivated by recent neural ordinary differential equations (ODE), we design an attention-inspired neural recursive ODE, named StockODE, which is more flexible for alleviating stock selection risks as well as promoting investment profits.

\subsection{Market Relation Learning}
According to Efficient Market Hypothesis (EMH)~\cite{fama1965behavior}, the stock movements are affected by several interactive facts, that is, the significant movements of a specific stock could result in the fluctuations of other stocks in the same industrial markets and the subsidiary stock price changes can also cause the upstream parent company's stock price movements. To this end, recent efforts turn to use the popular graph neural networks to incorporate the higher-order interactive correlations behind the stocks~\cite{ying2020time,cheng2021modeling,gao2021graph}. 
For example, GCN~\cite{chen2018incorporating} and GAT~\cite{hsu2021fingat} are two common but efficient solutions to aggregate the interaction impacts between the given stock and their relative stocks. Another solution turns to leverage the hypergraph for stock relation modeling to alleviate the information loss in 
traditional graph learning~\cite{sawhney2021stock}. Nevertheless, we argue that the domain-level interactions behind the stocks are under-explored. That is to say, our StockODE not only aggregates the given relationships among distinct stocks but also considers the potentially higher-order interactions upon pre-defined relation prior.

\section{Problem Statement}
\label{Problem Definition}

We aim at ranking a more profitable stock list for investors, assuming they have enough budget while lacking insights for investment decisions.
Let $\mathcal{S}=\{s_1,s_2,\cdots,s_i,\cdots,s_N\}$ be a set of $N$ stocks. For each stock $s_i$ on trading day $t_\tau$, it is associated with end-of-day data $\mathbf{x}_{i}^\tau (\in \mathbf{X}^\tau)$ including open, high, low, close prices and trading volume (i.e., $\mathbf{x}_{i}^\tau=[x_{i}^{o,\tau},x_{i}^{h,\tau},x_{i}^{l,\tau},x_{i}^{c,\tau},x_{i}^{v,\tau}]$ ). The indicator details are illustrated in Table~\ref{tab:des}. As the return ratio of a stock reveals the expected revenue of the stock~\cite{Feng2019TemporalRR,sawhney2021stock}, we also set a one-day return ratio $r_{i}^\tau=\frac{x_{i}^{c,\tau}-x_{i}^{c,\tau-1}}{x_{i}^{c,\tau-1}}$ for each stock $s_i$ on trading day $t_\tau$. 
Basically and formally, given historical stock data $\{\mathbf{X}^\tau\}_{T-w+1}^{T}$ with lookup window $w$, StockODE targets at predicting a ranking list of stocks $\mathbf{R}^{T+1}$ for the following trading day, where we have $\mathbf{R}^{T+1}=\{{R}_1^{T+1} > {R}_2^{T+1} \cdots > {R}_N^{T+1}\}$.  Specifically, for any two stocks $s_i$ and $s_j$ ($i \neq j$), we let ${R}_i^{T+1}>{R}_j^{T+1}$ if $r_{i}^{T+1}>r_{j}^{T+1}$.

\begin{table}[ht]
\centering
\fontsize{8.5}{9}\selectfont
\caption{The details of used stock indicators.}
\setlength{\tabcolsep}{0.5mm}{
\begin{tabular}{ll}
\toprule
\multicolumn{1}{l|}{Indicator}& \multicolumn{1}{c}{Description}  \\
\hline
\multicolumn{1}{l|}{Open}  & \multicolumn{1}{c}{The first price of the stock on a given trading day.} \\
\hline
\multicolumn{1}{l|}{Close} & \multicolumn{1}{c}{The final price of the stock on a given trading day.} \\
\hline
\multicolumn{1}{l|}{High}  & \multicolumn{1}{c}{The highest price of the stock on a given trading day.} \\
\hline
\multicolumn{1}{l|}{Low}  & \multicolumn{1}{c}{The lowest price of the stock on a given trading day.} \\
\hline
\multicolumn{1}{l|}{Volume}  & \multicolumn{1}{c} {\begin{tabular}[c]{@{}c@{}}The total amount of shares or contracts traded\\ for a particular security.\end{tabular}}\\
\bottomrule
\end{tabular}}
\label{tab:des}
\end{table}

\section{Methodology}
\label{Methodology}
\begin{figure*}[t]
    \centering
    \includegraphics[width=0.95\textwidth]{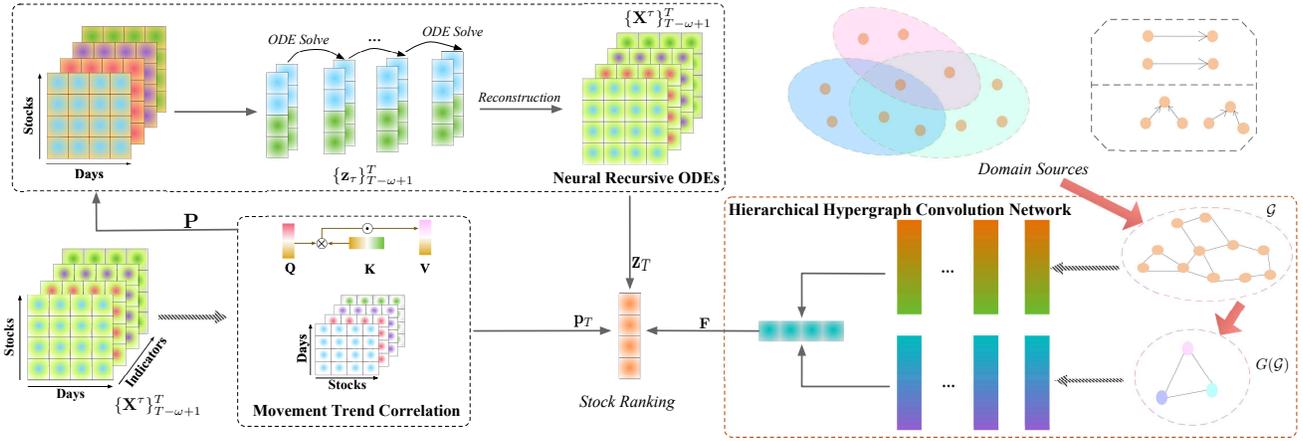}
    \caption{The framework of StockODE.}
    \label{fig:stockode}
\end{figure*}
In this section, we first introduce the skeleton of our proposed StockODE, followed by the details of each component. Finally, the optimization strategy and algorithm details are provided.
\subsection{Architecture Overview}
As illustrated in Fig.~\ref{fig:stockode}, it presents the overall framework of our \textbf{StockODE}, which mainly contains three components, i.e., \textit{Movement Trend Correlation}, \textit{Neural Recursive ODE}, and \textit{Hierarchical Hypergraph Convolution Network}. First, Movement Trend Correlation is to capture the explicit and implicit correlations among different stocks regarding the time-evolving return ratio, which results in a generation of aggregated results $\mathbf{P}$ by a standard self-attention neural network. Next, we devise a Neural Recursive ODE (NRODE) cell that operates $\mathbf{P}$ with the ODE solver in a recursive manner, whereby we can use it to generate the latent variables $\{\mathbf{z}_\tau\}_{T-\omega+1}^T$ via \textit{Variational Bayes}. Especially, NRODE coupling with an attention mechanism models a continuous dynamic system based on parameterizing the derivative of the latent state of each stock, instead of specifying the discrete sequence of hidden states' transformation. Subsequently, our Hierarchical Hypergraph Convolution Network (HHCN) turns to extract the intra- and inter-domain knowledge from multi-source domains, and then we fuse them to a unified representation $\mathbf{F}$. In the end, we use a simple dense layer to produce the predicted stock ranking, where the input contains the last hidden state $\mathbf{h}_T$, $\mathbf{z}_T$ and $\mathbf{F}$. We will elaborate on each component in the following part.

\subsection{Movement Trend Correlation}

\begin{figure}[ht]
    \centering
    \subfigure[Monday.]{\includegraphics[width=0.225\textwidth]{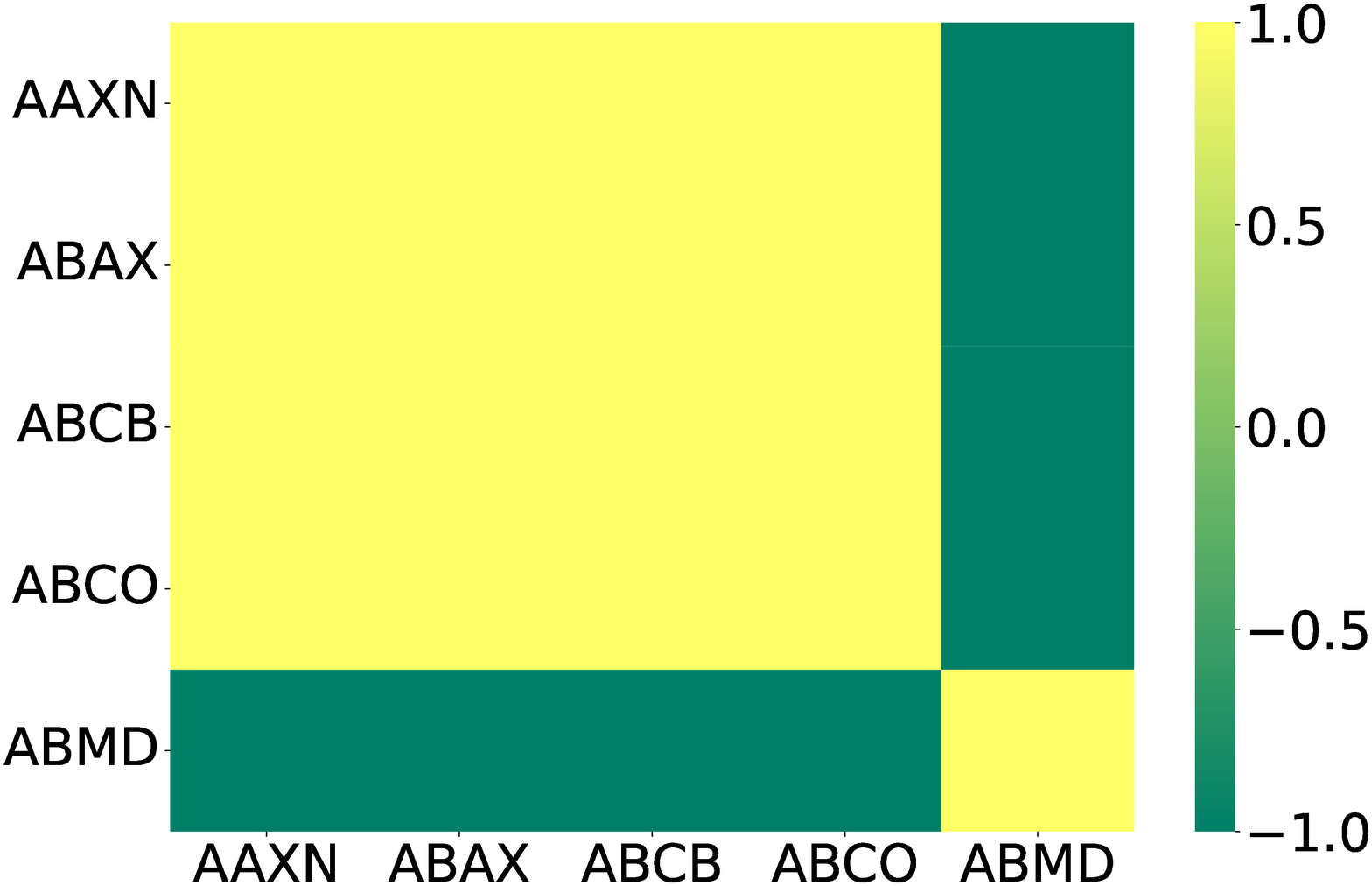}}
    \subfigure[Tuesday.]{\includegraphics[width=0.225\textwidth]{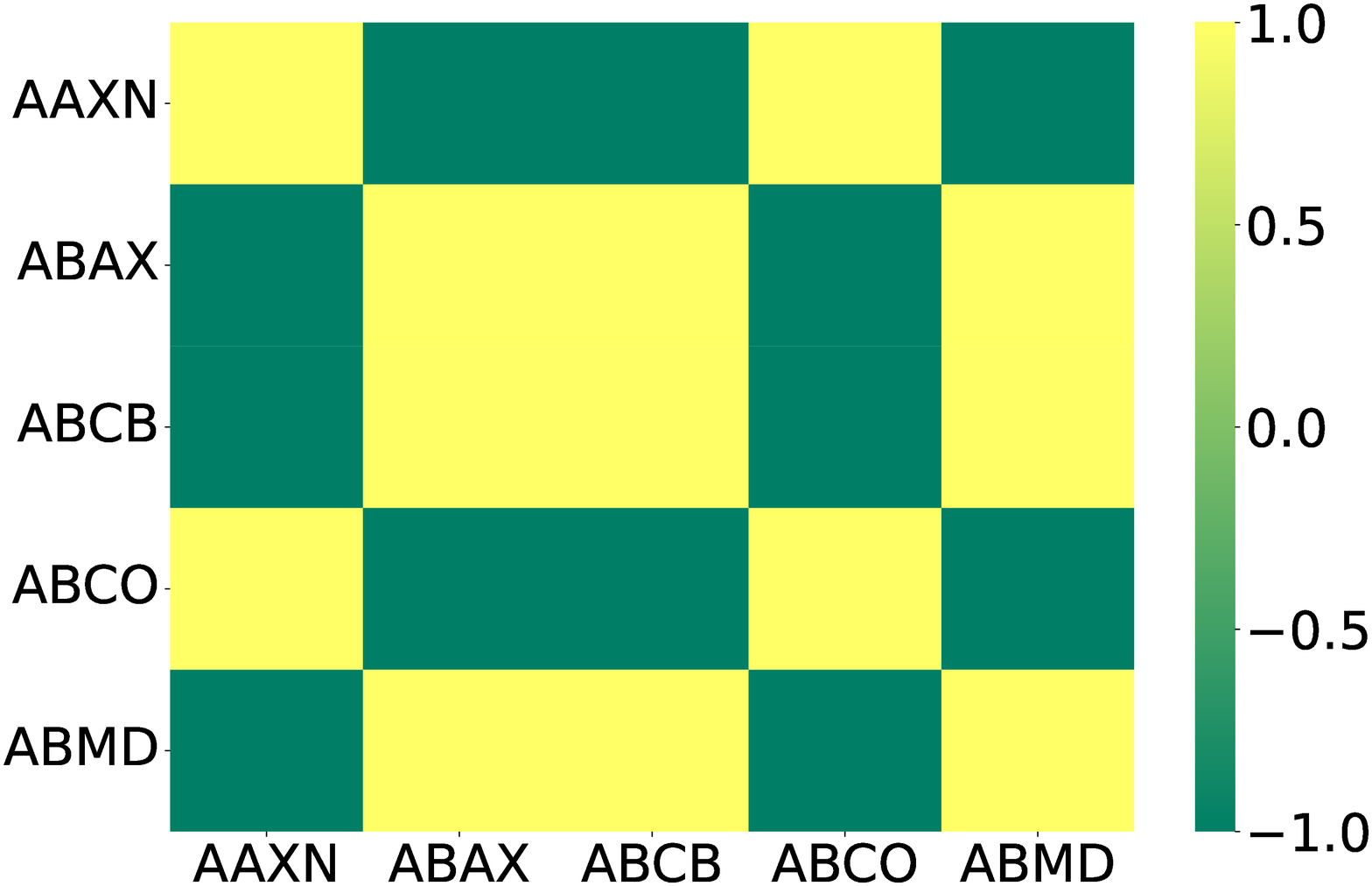}}
    \subfigure[Wednesday.]{\includegraphics[width=0.225\textwidth]{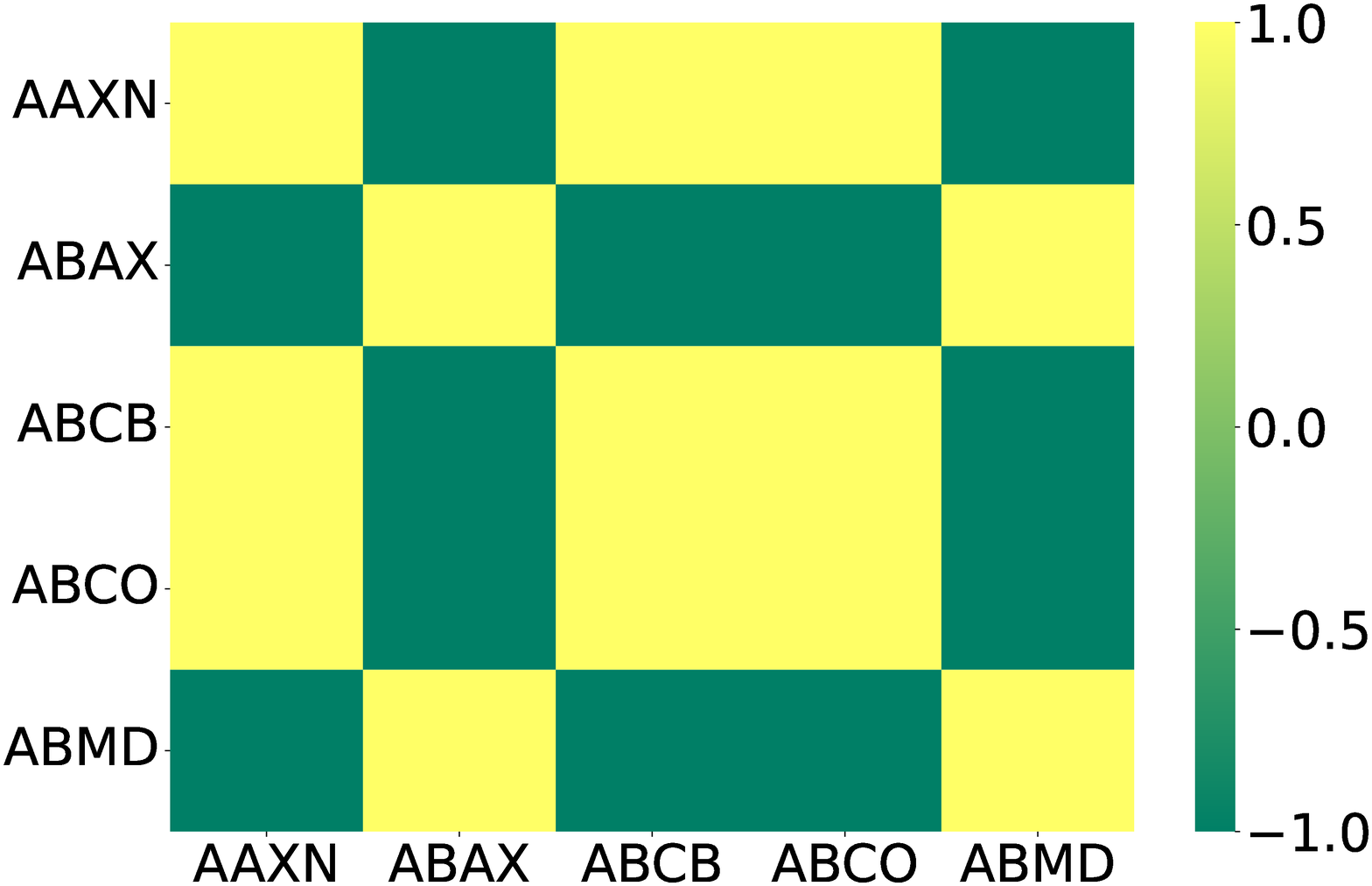}}
    \subfigure[Thursday.]{\includegraphics[width=0.225\textwidth]{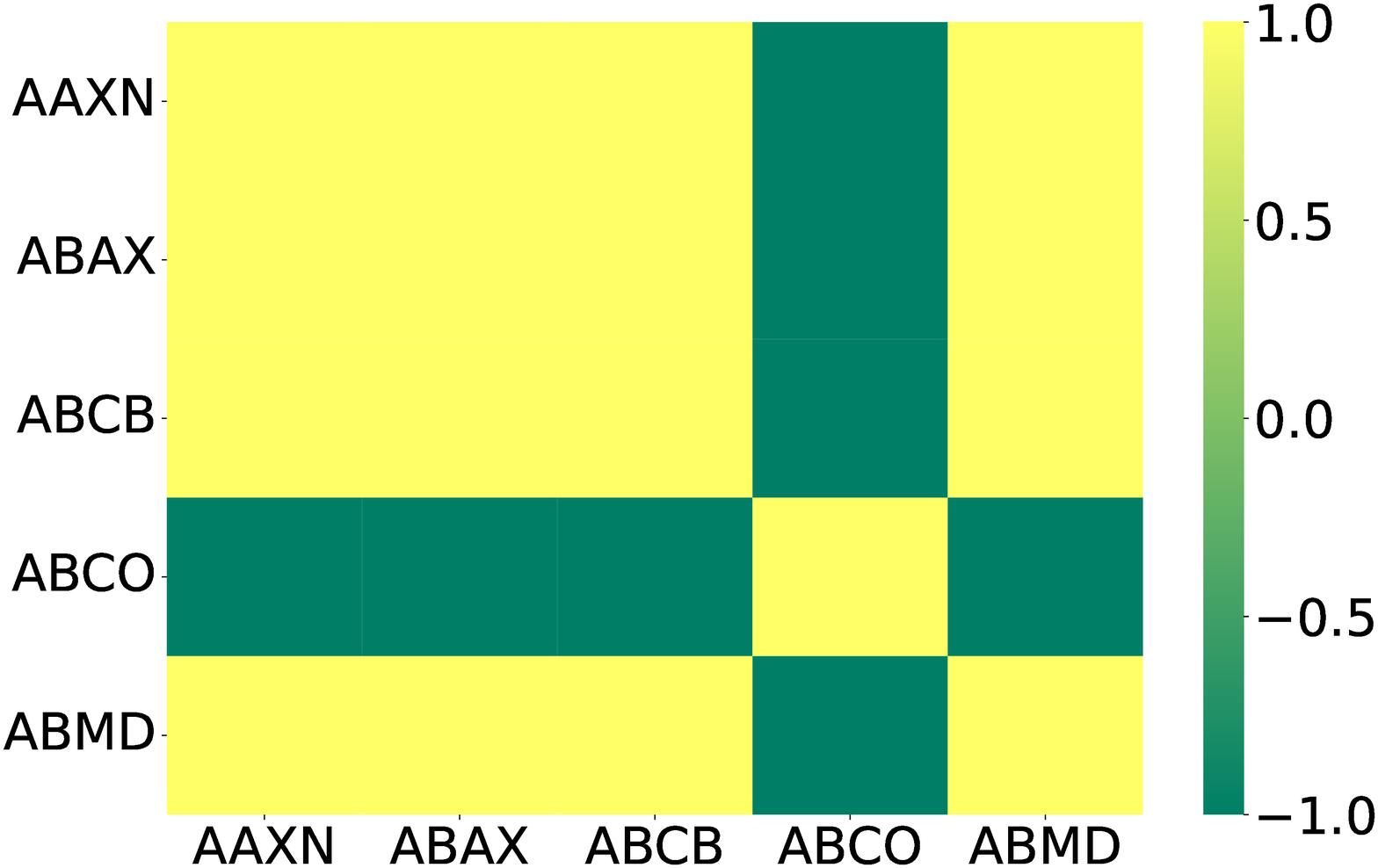}}
    \caption{A toy example of movement trend correlation from January 7, 2013 to January 10, 2013. Note that the yellow cell indicates the co-evolved relation while the green cell reflects the anti-evolved relation.}
    \label{fig:stockattention}
\end{figure}

Given stock set $\mathcal{S}$, we have their historical end-of-day data $\{\mathbf{X}^\tau\}_{T-w+1}^{T}$. We first transform $\{\mathbf{X}^\tau\}_{T-w+1}^{T}$ into a tensor form $[\mathbf{X}^{T-w+1},\cdots,\mathbf{X}^\tau,\cdots,\mathbf{X}^{T}]$, where $\mathbf{X}^\tau \in \mathbb{R}^{N\times d_e}$ preserves $N$ stocks' end-of-day indicators on trading day $t_\tau$, $d_e$ is the indicator number of each stock. For simplicity, we use $\mathbf{X} \in \mathbb{R}^{w\times N \times d_e}$ to denote such a tensor form. Since the 1-day return ratio only shows the daily return, we argue that considering the historical (middle) long-term earning impacts of each stock could help reduce the future investment risk. To this end, we also use return ratios of 5-, 10-, 20-, and 30-day moving averages to supplement stock features for addressing the influences of weekly and monthly trends. In practice, the return ratio can reveal the expected revenue of a specific stock. As shown in Fig.~\ref{fig:stockattention}, we randomly select five NASDAQ stocks and visualize their mutual correlations regarding the movement trends from Monday to Thursday in a given week. We can find that their mutual correlations, which vary with time, have drastically changed. For example, the stock \textit{AAXN} has an anti-evolved relationship with \textit{ABMD} from Monday to Wednesday, while \textit{AAXN} shows a co-evolved relationship with \textit{ABMD} on Thursday. we thus design an evolution-aware attention block containing an Explicit Correlation Aggregation and an Implicit Correlation Aggregation to expose the underlying correlations among different stocks regarding the time-evolving return ratios.  

\textbf{Explicit Correlation Aggregation.} We propose a correlation tensor $ \boldsymbol{\Upsilon}$ ($\in \mathbb{R}^{w\times N \times N}$) regarding the return ratio to expose the co-evolved and anti-evolved relationships among different stocks. Specifically, the 1-day return ratio indicates the difference between the current trading day and the previous trading day. For a given trading day $t_\tau$, we thus define the following formulation to obtain the correlation coefficient among different stocks w.r.t the trading day $t_\tau$'s return ratio:
\begin{footnotesize}
\begin{align}
\label{eq:ex-corre}
    \centering
    &\boldsymbol{\Upsilon}^\tau=
    &\textbf{sign}
    \begin{bmatrix}
    r_{1}^\tau &r_{2}^\tau&\cdots&r_{N}^\tau\\
    r_{1}^\tau &r_{2}^\tau&\cdots&r_{N}^\tau\\ \vdots&\vdots&\ddots&\vdots\\
    r_{1}^\tau &r_{2}^\tau&\cdots&r_{N}^\tau\\
    \end{bmatrix}\otimes\textbf{sign}
    \begin{bmatrix}
    r_{1}^\tau &r_{2}^\tau&\cdots&r_{N}^\tau\\
    r_{1}^\tau &r_{2}^\tau&\cdots&r_{N}^\tau\\ \vdots&\vdots&\ddots&\vdots\\
    r_{1}^\tau &r_{2}^\tau&\cdots&r_{N}^\tau\\
    \end{bmatrix}^\top,
\end{align}
\end{footnotesize}
where $\textbf{sign}$ is a Sign function and $\otimes$ refers to element-wise multiplication. $\boldsymbol{\Upsilon}^\tau$ is able to indicate either a positive or negative correlation between different stocks in terms of their return ratios. As such, we can eventually formulate a correlation tensor $\boldsymbol{\Upsilon}$ to describe the return ratio-aware relations on each trading day, i.e., $\boldsymbol{\Upsilon}=[\boldsymbol{\Upsilon}^1,\boldsymbol{\Upsilon}^2,\cdots,\boldsymbol{\Upsilon}^\omega]$.

Next, we operate a one-layer graph convolutional layer to aggregate explicit correlations among different stocks, which can be summarized as follows:
\begin{align}
\label{eq:input}
&\mathbf{H}={\boldsymbol{\Upsilon}}\mathbf{X} \mathbf{W}_\Upsilon,\\\nonumber
&\mathbf{H}=\mathbf{H}+\mathbf{X}\mathbf{W}_X,
\end{align}
where $\mathbf{W}_\Upsilon (\in \mathbb{R}^{d_e\times d})$ and $\mathbf{W}_X (\in \mathbb{R}^{d_e\times d})$ are trainable parameters.

\textbf{Implicit Correlation Aggregation.} Inspired by~\cite{vaswani2017attention}, we employ a self-attention layer to capture the attentive stock correlations w.r.t $\mathbf{H}$. Note that this self-attention aims to evaluate the correlation scores among the stocks instead of capturing the temporal dependencies of the stock itself. Specifically, it first computes the query $\mathbf{Q}$, key $\mathbf{K}$ and value $\mathbf{V}$, and then uses dot-product attention with an activation function to generate the latent states $\mathbf{H}'$ ($\in \mathbb{R}^{w\times N \times d}$). Thus, the process can be written as:
\begin{equation}
\label{eq:self-att}
\begin{aligned}
&\mathbf{H}'=\operatorname{ReLU}\left(\operatorname{Norm}\left(\frac{\boldsymbol{\Omega} \cdot \mathbf{H} \boldsymbol{W}_{V}}{\sqrt{d}}\right)\right), \\
&\boldsymbol{\Omega}=\operatorname{softmax}\left(\mathbf{H} \boldsymbol{W}_{Q} \cdot \left(\mathbf{H} \boldsymbol{W}_{K}\right)^{\top}\right),\\
\end{aligned}
\end{equation}
where $\mathbf{H} \boldsymbol{W}_{Q}$, $\mathbf{H} \boldsymbol{W}_{K}$ and $\mathbf{H} \boldsymbol{W}_{V}$ refer to $\mathbf{Q}$, $\mathbf{K}$ and $\mathbf{V}$ in the self-attention neural network, respectively. $\operatorname{Norm}$ denotes the Layer Normalization operation
for fast and stable training. And the self-attention score matrix $\boldsymbol{\Omega}$ is normalized by a softmax function. Notably, $\boldsymbol{W}_{Q}, \boldsymbol{W}_{K}, \boldsymbol{W}_{V} \in \mathbb{R}^{d\times d}$ are trainable matrices and $d$ refers to the size of dimensionality. 
Now we reuse the transpose operation to shift $\mathbf{H}'$ to temporal perspective:
\begin{equation}
\label{eq:P}
\mathbf{P}=\operatorname{tran}(\mathbf{H}'),
\end{equation}
where $\mathbf{P} \in \mathbb{R}^{N \times w \times d}$. We will take $\mathbf{P}$ as the input of the following Neural Recursive ODEs.

\subsection{Neural Recursive ODEs}
We now introduce our devised Neural Recursive ODE (NRODE) for multivariate stock movement learning. Prior to that, we first go over the technical specifics of current neural ODEs and show how our NRODE is inspired and what it contributes.

\textbf{Neural ODE.} Recent neural ODE enables generating continuous observations by giving the initial condition~\cite{chen2018neural}, which stimulates us to develop the neural ODE-based module to tackle the discrete stock indicators (observations). Neural ODEs (NODEs) intuitively build the infinite-steps hidden state update in neural networks for bridging the gap between discrete neural networks and continuous dynamic systems~\cite{chen2018neural,jhin2021ace}. Specifically, NODEs, starting from the input hidden layer $\mathbf{h}(t_0)$, parameterize the continuous dynamics of hidden units using an ODE specified by a neural network:
\begin{align}
\label{eq:ode}
&\frac{d \mathbf{h}(t)}{d t}=f(\mathbf{h}(t), t;\boldsymbol{\psi}), \text{where}\\\nonumber
&\mathbf{h}\left(t_{1}\right)=\mathbf{h}\left(t_{0}\right)+\int_{t_{0}}^{t_{1}} f\left(\mathbf{h}(t), t ; \boldsymbol{\psi}\right) d t.
\end{align}
Notably, ODE function $f\left(\boldsymbol{h}(t), t ; \boldsymbol{\psi}\right)$ is a neural network parameterized by $\boldsymbol{\psi}$. And the above process also can be rewritten as:
\begin{align}
\label{eq:ode-s}
&\mathbf{h}\left(t_{1}\right)=\text{ODESolve}(f_{\boldsymbol{\psi}},\mathbf{h}\left(t_{0}\right),(t_0,t_1))
\end{align}
As such, the most significant benefit of ODE is that we can obtain the results of dynamic hidden representation at any time (e.g., $t_1$) when we have received the initial state at $t_0$ ($t_0<t_1$).

\begin{figure}[ht]
    \centering
    \includegraphics[width=0.5\textwidth]{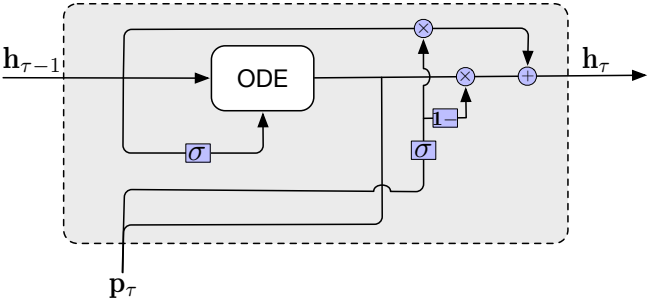}
    \caption{A circuit diagram of NRODE cell. Note that $\sigma$ is the `sigmoid' operation, $\otimes$ is the element-wise multiplication, and $\oplus$ refers to the sum operation.}
    \label{fig:NRODE}
\end{figure}

As one of the popular neural ODEs, the ODE-GRUs~\cite{rubanova2019latent} could provide us with a new paradigm for stock movement learning, which can be summarized as: 
\begin{align}
\label{eq:ode-rnn}
&\mathbf{h}'_{i}=\text{ODESolve}(f_{\boldsymbol{\psi}},\mathbf{h}_{i-1},(t_{i-1},t_{i})),\\
&\mathbf{h}_i=\text{GRUCell}(\delta_i,\mathbf{h}'_i),
\end{align}
where $\mathbf{\delta}_i$ denotes the input feature and $\mathbf{h}_i$ is the hidden state of current input $\delta_i$.
Nevertheless, the trend of stock indicators such as price is different from the representative time series modeling problems as the former usually does not has strong regularity which is dependent on time while the latter usually follows obvious transitional patterns/regularities (e.g., human mobility~\cite{liang2021modeling}, urban flow~\cite{zhou2020enhancing}, residential electricity consumption~\cite{dong2016hybrid}, and etc).
The reason is stock movements in the future are usually disturbed by multiple external factors (e.g., public opinion, stimulus policy, investor psychology, and etc.). As a result, we conjecture that the unobserved or unconsidered uncertainty of the stock market could also bring the risk of stock selection. To this end, we devise a lightweight recursive model--Neural Recursive ODE (NRODE)--which enables co-evolving the stock movements recursively with the attention mechanism. As shown in Fig.~\ref{fig:NRODE}, our NRODE discards the complex gate operations in vanilla RNN (e.g., LSTM and GRU). Instead, we propose an attention-inspired ODE block to generate the candidate of continuous hidden state whereafter incorporating the current discrete observation to formulate the current hidden state. In particular, we encode these hidden states into latent codes via Variational Bayes to account for the uncertainty in stock movements.

\textbf{NRODE.} Given the initial hidden states of stock movements $\mathbf{P}$, we first adopt a linear transformation layer to tackle each trading day data $\mathbf{p}_\tau$ ($\in \mathbf{P}$) and set the obtained results as the input of NRODE:
\begin{align}
\label{eq:in-nrode}
        &\mathbf{v}_\tau=\mathbf{p}_{\tau}\boldsymbol{W}_{v}+\boldsymbol{b}_{v},
\end{align}
where $\boldsymbol{W}_{v} \in \mathbb{R}^{d\times d }$ and $\boldsymbol{b}_{v} \in \mathbb{R}^{d }$ are learnable parameters. For each $\mathbf{v}_\tau$ along with time $t_\tau$, we have:
\begin{align}
\label{eq:odeblock0}
    &\mathbf{h}'_{\tau}=\text{ODESolve}(f_{\boldsymbol{\psi}},(\mathbf{h}_{\tau-1},a(\mathbf{h}_{\tau-1})),(t_{\tau-1},t_{\tau})),\\
    &\mathbf{h}_{\tau}''=[\mathbf{v}_\tau,\mathbf{h}'_{\tau}]\boldsymbol{W}_h+\boldsymbol{b}_h,\\
    &\mathbf{h}_{\tau}=(\mathbf{1}-\mathbf{I}(\mathbf{v}_\tau))\mathbf{h}_{\tau}''+\mathbf{I}(\mathbf{v}_\tau)\mathbf{h}_{\tau-1}
\end{align}
where $\mathbf{h}_0$ is set as $\mathbf{0
}$ and $\mathbf{I}(\mathbf{v}_\tau)$ denotes the update gate. $\mathbf{h}_\tau$ is the updated hidden state and $f_{\boldsymbol{\psi}}$ is the differentiable network parameterized by $\boldsymbol{\psi}$. Herein, we choose the Euler solution as the numerical solver in our StockODE instead of the adjoint method due to its numerical instability of backward ODE solve~\cite{chen2018neural}. Inspired by previous works~\cite{liang2021,zhou2021forecasting}, we also employ three-layer multilayer perceptrons (MLPs) as the ODE function where each layer has $d$ units.
And we notice that Eq.~(\ref{eq:odeblock0}) can be theoretically described as follows:
\begin{align}
&\boldsymbol{h}'\left(t_{\tau}\right)=\boldsymbol{h}\left(t_{\tau-1}\right)+\int_{t_{\tau-1}}^{t_{\tau}} f\left(\boldsymbol{h}(t), \boldsymbol{a}(t), t ; {\boldsymbol{\psi}}\right) d t,
\end{align}
where $\mathbf{a}(t)$ has the same dimension as $\mathbf{h}(t)$ and it aims to regularize the contribution of each value in $\mathbf{h}(t)$. In particular, the ODE function $f$ can be further described as:

\begin{align}
f\left(\boldsymbol{h}(t), \boldsymbol{a}(t), t ; {\boldsymbol{\psi}}\right) &=f\left(g(\boldsymbol{h}(t),\boldsymbol{a}(t)), t ; {\boldsymbol{\psi}}\right),
\end{align}
where $g(\cdot,\cdot)$ is an attention-integrated layer, defined as:
\begin{equation}
    \label{eq:att-g-func}
    g(\mathbf{h}(t),a(\mathbf{h}(t)))=\mathbf{h}(t)\otimes \text{sigmoid}(a(\mathbf{h}(t))),
\end{equation}
where $\otimes$ is element-wise multiplication. Notably, the input of attention state $\mathbf{a}(t)$ is initialized by $\mathbf{h}(t)$:
\begin{align}
    \label{eq:attention}
    \boldsymbol{a}(t)=\mathbf{h}(t)\boldsymbol{W}_{a}+\boldsymbol{b}_{a},
\end{align}
where $\boldsymbol{W}_{a}\in \mathbb{R}^{d \times d}$ and $\boldsymbol{b}_{a} \in \mathbb{R}^{d}$ denote the trainable parameters, respectively.

\textbf{Posterior Approximation.} To consider the uncertainty/stochasticity that is hard to observe in stock movements, we turn to operate each hidden state into a latent space based on Variational Bayes (VB)~\cite{kingma2013auto}. Specifically, we can obtain a sequence of latent variables $\mathbf{Z}=\{\mathbf{z}_\tau\}_{T-w+1}^{T}$, which enables tackling the uncertain factors in stock movements. Each latent variable $\mathbf{z}_{\tau}$ can be formulated as follows:
\begin{align}
    &\mathbf{z}_{\tau}=\boldsymbol{h}_{\tau}+\boldsymbol{\sigma}_{\tau}\boldsymbol{\epsilon}, \\
    &[\boldsymbol{\mu}_\tau,\boldsymbol{\sigma}_{i}]=\boldsymbol{h}_{\tau}\boldsymbol{W}_z+\boldsymbol{b}_z.
\end{align}
Herein, $\boldsymbol{\epsilon}$ is sampled from Gaussian noise, i.e., $\boldsymbol{\epsilon} \sim \mathcal{N}(\mathbf{0},\mathbf{1})$. 
The general pipeline of NRODE operated in a recursive manner is summarized in Algorithm ~\ref{algorithm-att-ode}.
\begin{algorithm}[t]
\label{algorithm-att-ode}
	\caption{the workflow of NRODE.}
	\SetKwInput{KwInput}{Input}        
	\SetKwInput{KwOutput}{Output} 
	\KwInput{$\{\mathbf{p}_\tau \}_{\tau=T-\omega+1}^T$}
	let $\mathbf{h}_0=\mathbf{0}$;\\
	\For{$\tau=T-\omega+1$; $\tau=\tau+1$; $\tau \leq T$}{
	$\mathbf{v}_\tau=\mathbf{p}_{\tau}\boldsymbol{W}_{v}+\boldsymbol{b}_{v}$;\\
	$a(\mathbf{h}_{\tau-1})=\mathbf{h}_{\tau-1}\mathbf{W}_a+\mathbf{b}_a$;\\
	$\mathbf{h}'_\tau=\text{ODESolve}(f_{\boldsymbol{\psi}},(\mathbf{h}_{\tau-1},a(\mathbf{h}_{\tau-1})),(t_{\tau-1},t_\tau))$;\\
	$\mathbf{h}''_{\tau}=[\mathbf{v}_\tau,\mathbf{h}'_{\tau}]\boldsymbol{W}_\mu+\boldsymbol{b}_\mu$;\\
	$\mathbf{h}_{\tau}=\mathbf{I}(\mathbf{v}_i)\mathbf{h}_{\tau}''+(\mathbf{1}-\mathbf{I}(\mathbf{v}_\tau))\mathbf{h}_{\tau-1}$;\\
	$\mathbf{z}_{\tau}=\boldsymbol{h}_{\tau}+\boldsymbol{\sigma}_{\tau}\boldsymbol{\epsilon};$
	}
	\KwOut{$\{\mathbf{z}_\tau\}_{\tau=T-\omega+1}^T$}
\end{algorithm}

\subsection{Hierarchical Hypergraph Convolution Network}

As the popular hypergraph learning enables modeling complex high-order relations, we construct a hierarchical hypergraph convolution network (HHCN) to describe the domain-aware dependencies regarding the stocks from two views, i.e., intra-domain and inter-domain. In short, we first construct a hypergraph $\mathcal{G}$ from various kinds of domain interactions to extract the intra-domain Knowledge. Then, we build a meta-hypergraph $G(\mathcal{G})$ based on the hypergraph $\mathcal{G}$ to address the domain-level interactions for the acquisition of inter-domain knowledge. Below are the details.

\subsubsection{Intra-domain Knowledge} Let $\mathcal{G}=(\mathcal{S},\mathcal{E})$ denote a hypergraph, where $\mathcal{S}$ is a set containing $N$ stocks ( i.e., $s_i\in \mathcal{S}, N=|\mathcal{S}|$) and $\mathcal{E}$ represents a set of hyperedges. Each hyperedge $e_j \in \mathcal{E}$ contains two or more stocks, reflecting an intra-domain fact, e.g., they are from the same industry. Note that the real-world stock relations (e.g., supplier-consumer relation and ownership relation) are derived from the third-party open data $\mathcal{D}_r$ such as Wiki~\cite{vrandevcic2014wikidata}. 
Additionally, we assign a positive weight $\mathbf{w}_{jj}$ to each hyperedge $e_j$, and finally formulate a diagonal matrix $\boldsymbol{\Psi} \in \mathbb{R}^{|\mathcal{E}|\times |\mathcal{E}|}$. Similar to previous works~\cite{xia2021self,sawhney2021stock}, we set $\boldsymbol{\Psi}$ as the identity matrix, which indicates equal weights for all hyperedges. Actually, the hypergraph $\mathcal{G}$ can be equivalently denoted by an incidence matrix $\mathbf{M} \in \mathbb{R}^{N\times |\mathcal{E}|}$ where $\mathbf{M}_{ij}=1$ if the hyperedge $e_j \in \mathcal{E}$ contains a stock $s_i \in \mathcal{S}$, otherwise 0. Correspondingly, we can respectively define $\mathbf{D}_{ii}$ and $\mathbf{O}_{jj}$ as the degree of a stock $s_i$ and a hyperedge $e_j$, where $\mathbf{D}_{ii}=\sum_{e_j\in \mathcal{E}}\boldsymbol{\Psi}_{jj}\mathbf{M}_{ij}$ and $\mathbf{O}_{jj}=\sum_{i=0}^{N}\mathbf{M}_{ij}$. Finally, we obtain the degree matrices $\mathbf{D}\in \mathbb{R}^{N\times N}$ and $\mathbf{O} \in \mathbb{R}^{|\mathcal{E}|\times |\mathcal{E}|}$, where $\mathbf{D}$ and $\mathbf{O}$ are diagonal matrices. Following~\cite{xia2021self}, we use multi-layer hypergraph convolutions to aggregate the feature information, which inherently comprises two-stage knowledge passing, followed by stock-hyperedge and hyperedge-stock. For instance, the $l$-th layer can be represented as:
\begin{equation}
\label{eq:intra-graph}
\mathbf{U}^{(l+1)}=\mathbf{D}^{-1} \mathbf{M} \mathbf{\boldsymbol{\Psi} O}^{-1} \mathbf{M}^{\intercal} \mathbf{U}^{(l)}\mathbf{W}_U^{(l)},
\end{equation}
where $\mathbf{W}_U^{(l)}$ is a trainable matrix.
Finally, the intra-domain knowledge among the stock will be successfully incorporated in $\mathbf{U}^{(L)} \in \mathbb{R}^{N \times d'}$ after passing through $L$ hypergraph layers. 

\subsubsection{Inter-domain Knowledge}
Inter-domain knowledge reflects the domain-level interaction behind stocks. Given the hypergraph $\mathcal{G}=(\mathcal{S},\mathcal{E})$, we present a meta-hypergraph $G(\mathcal{G})$ to explore the inter-domain knowledge where each meta-node $c_i$ in $G(\mathcal{G})$ is a hyperedge in $\mathcal{G}$. Actually, meta-hypergraph $G(\mathcal{G})$ depicts the connectivity of hyperedges in $\mathcal{G}$. That is to say, the meta-edge between meta-node $c_i$ and meta-node $c_j$ refers to their corresponding hyperedges in $\mathcal{G}$ have at least one common stock. In a nutshell, $G(\mathcal{G})(C,E)$ contains a meta-node set $C=\{c_e:c_e\in \mathcal{E}\}$ and a meta-edge set $E=\{(c_{e_m},c_{e_n}):c_{e_m},c_{e_n} \in \mathcal{E}, \|e_m \cap e_n\| \geq 1 \}$. In addition, each edge $(c_{e_m},c_{e_n})$ is associated with a weight $\boldsymbol{\Omega}_{mn}$ calculated by $\frac{|e_m\cap e_n|}{|e_m\cup e_n|}$. As such, we can obtain a weight matrix $\boldsymbol{\Omega}$ for $G(\mathcal{G})$ and regard it as the incidence matrix. Now we can operate the simple graph convolutional networks for information distillation, where each layer can be defined as:
\begin{equation}
\label{eq:inter-graph}
\mathbf{B}^{(l+1)}=\hat{\mathbf{D}}^{-1} \hat{\boldsymbol{\Omega}} \mathbf{B}^{(l)} \mathbf{W}^{(l)},
\end{equation}
where $\hat{\mathbf{D}}$ is diagonal degree matrix of $\hat{\boldsymbol{\Omega}}$, $\hat{\boldsymbol{\Omega}}=\boldsymbol{\Omega}+\mathbf{I}$, and $\mathbf{W}_{\mathbf{B}}^{(l)} \in \mathbb{R}^{d'\times d'}$ is a learnable matrix. Herein, $\mathbf{I}$ is an identity matrix. In the end, we obtain the the final inter-domain knowledge $\mathbf{B}^{(L)} \in \mathbb{R}^{|\mathcal{E}|\times d'}$ after $L$ convolutional operations.
\subsubsection{Knowledge Interaction}
To incorporate both intra- and inter- domain knowledge for stock recommendation, we design an interactive operation to formulate the fused knowledge $\mathbf{F}$, which is denoted as:
\begin{equation}
\label{eq:graph-fus}
    \mathbf{F}=\mathbf{U}^{(L)}{\mathbf{B}^{(L)}}^\intercal \mathbf{W}_F,
\end{equation}
where $\mathbf{W}_F \in \mathbb{R}^{|\mathcal{E}| \times d}$ is a learnable matrix. We will incorporate the fused knowledge for downstream stock selection.

\subsection{Optimization}
We now introduce how to incorporate a variety of extracted features into a unified formulation for task prediction. And then we illustrate the training details including the final optimization object and algorithmic aspect.

\textbf{Knowledge Fusion.}
As we obtain the temporal correlation information and domain knowledge as well as learned latent variables $\mathbf{z_T}$, we concatenate them and employ a simple dense neural network as a fusion layer to generate the predicted return ratios $\hat{\boldsymbol{r}}^{T+1}$, which can be formulated as:
\begin{align}
\label{obj-fusion}
    \hat{\boldsymbol{r}}^{T+1}=\text{LeakyReLU}([\mathbf{p}_T;\mathbf{z}_T;\mathbf{F}]\mathbf{W}_r+\mathbf{b}_r),
\end{align}
where $\text{LeakyReLU}$ is the activation function and $;$ refers to the concatenation operation. In addition, $\mathbf{W}_r \in \mathbb{R}^{d\times d}$ and $\mathbf{b}_r \in \mathbb{R}^{d}$ are trainable parameters. Finally, we can produce the ranked stock list according to the predicted results.

\begin{algorithm}[t]
	\caption{StockODE Training.}
	\label{algorithm-stockode}
	\SetKwInput{KwInput}{Input}        
	\SetKwInput{KwOutput}{Output} 
	\KwInput{stock set $\mathcal{S}$,
	historical end-of-day data $\{\mathbf{X}\}_{T-\omega+1}^T$, and stock relation data $\mathcal{D}_r$.}
	\tcc{Preparation}
	initialize the parameters $\Theta$ in StockODE;\\
	Formulate the correlation tensor $\boldsymbol{\Upsilon}$ via Eq.(\ref{eq:ex-corre});\\
	Construct hypergraph $\mathcal{G}$ from $\mathcal{D}_r$;\\
	Construct meta-hypergraph $G(\mathcal{G})$ based on $\mathcal{G}$;\\
	\For{epoch i=0; i=i+1; $i<N$}{
			\tcc{\textit{Movement Trend Correlation}}
	Obtain explicit correlations $\mathbf{H}$ via Eq.(\ref{eq:input});\\
	Generate $\mathbf{P}$ via Eq.(\ref{eq:self-att}) and Eq.(\ref{eq:P});\\
		\tcc{\textit{NRODE}}
			Operate the Algorithm~\ref{algorithm-att-ode} to obtain $\mathbf{z}_T$; \\
			\tcc{\textit{HHCN}}
			Explore the intra-domain knowledge $\mathbf{U}^{(L)}$ via Eq.(\ref{eq:intra-graph});\\
			Explore the inter-domain knowledge $\mathbf{B}^{(L)}$ via Eq.(\ref{eq:inter-graph});\\
			Obtain fused graph knowledge $\mathbf{F}$ via Eq.(\ref{eq:graph-fus});\\
			\tcc{Optimization}
            Generate the stock ranking $\hat{\boldsymbol{R}}^{T+1}$ via Eq.(\ref{obj-fusion}); \\
			Compute the training loss according to Eq.(\ref{eq:opt});\\
		Update the learnable parameters;\\
	}
	\KwOut{the optimal Model $\Theta^*$.}
\end{algorithm}

\textbf{ELBO.}
StockODE using latent variables also should optimize the variational divergence by maximizing the following evidence lower bound (ELBO):
\begin{align}
    \label{eq:elbo}
    \mathcal{J}_{ELBO}(\theta,\phi)=\mathbb{E}_{q_\phi} \log [p_\theta(\mathbf{X}|\mathbf{Z})]+\mathbb{E}_{q_\phi} \log [p_\theta(\mathbf{Z})]\\ \nonumber
    -\mathbb{E}_{q_\phi} \log [q_\phi(\mathbf{Z}|\mathbf{X})].
\end{align}
Wherein, the first term is the reconstruction likelihood and the last two terms are the KL-divergence of prior assumption (i.e., Gaussian distribution) and the variational posterior distribution. We regard our NRODE as the cognition network $q_\phi$ while using another NRODE as the generative network $p_\theta$.

\textbf{Multi-task Learning.}
In order to minimize the loss between the predicted return ratios and ground truth while keeping the ranked stocks with higher returns for investment, StockODE which is operated in a multi-task learning manner uses a point-wise regression and pairwise ranking-aware loss function to optimize the trainable parameters. The reason is that our goal, which is to discover more profitable stocks, cannot be reflected by merely applying the point-wise regression function (e.g., mean squared error). Thus, we also employ the pairwise ranking-aware loss function to optimize the ranking of stocks~\cite{hsu2021fingat}. In addition, we also should maximize the ELBO to explore the intricate distribution behind the stock movement patterns. In the end, the optimization object can be summarized as:
\begin{align}
\label{eq:opt}
\mathcal{L}(\Theta)&=\left\|\hat{\boldsymbol{r}}^{T+1}-\boldsymbol{r}^{T+1}\right\|^{2}-\beta \mathcal{J}_ {ELBO}\\\nonumber
+& \sum_{i=0}^{|\mathcal{S}|} \sum_{j=0}^{|\mathcal{S}|} \max \left(0,-\left(\hat{\boldsymbol{r}}_{i}^{T+1}-\hat{\boldsymbol{r}}_{j}^{T+1}\right)\left(\boldsymbol{r}_{i}^{T+1}-\boldsymbol{r}_{j}^{T+1}\right)\right),
\end{align}
where the first term is point-wise regression loss, the second term is ELBO, and the last one is pairwise ranking loss regarding return ratio. Herein, $\Theta$ refers to the trainable parameters in StockODE. $\beta$ is a trade-off factor. The general training pipeline of StockODE is presented in Algorithm~\ref{algorithm-stockode}.

\section{Experiments}
\label{Experiments}
We now present the details of experimental evaluations to verify the effectiveness of our proposed StockODE. Specifically, we first introduce experimental settings including datasets, metrics, and baselines, followed by implementation details. Next, we compare the experimental results of StockODE with several state-of-the-art baselines. Finally, the ablation study, continuity analysis, and sensitivity analysis are presented.
\subsection{Experimental Setup}
\subsubsection{Datasets.} To verify the performance of our proposed method, we conduct the experiments on two real-world datasets collected from NASDAQ and NYSE market~\cite{Feng2019TemporalRR}. Specifically, the collected stocks from the NASDAQ and NYSE have transaction records between January 2, 2013 and November 8, 2017, where each selected stock has been traded on more than 98\% of trading days and has never been traded at less than \$ 5 per share. In the end, we respectively obtain 1,026 and 1,737 stocks for our experiments. Notably, each dataset contains three types of data, including historical indicator data, sector-industry relations, and Wiki relations between their companies such as supplier-consumer relations and ownership relations. Specifically, we follow previous work~\cite{sawhney2021stock} to formulate three types of predefined hyperedges that describe the first-order and second-order relations stemming from Wikidata besides sector-industry relations. The details are shown in Table~\ref{tab:data}. We choose the first 756 days' stock trades for training, the next 252 days' stock trades for validating, and the remaining for testing.

\begin{table}[ht]
\caption{Statistics of stock trading data.}
\centering
\fontsize{7.5}{7.5}\selectfont
\setlength{\tabcolsep}{1.0mm}{
\begin{tabular}{l|c|c}
 \toprule
  \textbf{Description}& \textbf{NASDAQ} & \textbf{NYSE} \\
 \hline
 Period   & 01/02/2013-12/08/2017    &01/02/2013-12/08/2017 \\
 \hline
 Days(Train:Val:Test) &  756:252:237  & 756:252:237   \\
  \hline
 Stocks &1026 & 1737\\
  \hline
 Industries Types &112 & 130\\
 \hline
 Wiki Types &  42  & 32\\
  \hline
 Hyperedges &  1142  &1979\\
\bottomrule
\end{tabular}}

\label{tab:data}
\end{table}
\subsubsection{Baselines.} We compare our StockODE with representative conventional and neural network-based methods. The details are:
\begin{itemize}
    \item \textbf{ARIMA}~\cite{ariyo2014stock} first applies a differencing transformation to a given time series, then uses autoregressive (AR) and moving average (MA) for stock ranking.
    \item \textbf{LSTM}~\cite{chen2015lstm} operates a simple LSTM neural network for stock return prediction, where multiple indicators regarding the stocks are considered, including trading volume, open price, close price, and etc.
    \item \textbf{GRU}~\cite{shen2018deep} is another widely used RNN model. Similar to LSTM~\cite{chen2015lstm}, we use it to model the stock movements.
        
    \item \textbf{DA-RNN}~\cite{qin2017dual} designs a dual-stage attention-based RNN for historical stock movement modeling.
    \item \textbf{CNNpred}~\cite{hoseinzade2019cnnpred} aggregates several market factors in a CNN-based framework for feature extraction and financial markets behavior modeling.
    \item \textbf{StockNet}~\cite{xu2018stock} is a deep generative model that uses the variational Bayes to exploit stock price signals as well as other textual information.
    \item \textbf{LSTM+GCN}~\cite{chen2018incorporating} uses a standard LSTM model to tackle historical stock prices and adopts the simple GCN to explore the stock correlations.
    \item \textbf{RSR}~\cite{Feng2019TemporalRR} is a temporal graph-based method, which considers the temporal evolution and relation network of stocks.
    \item \textbf{HATR}~\cite{wang2021} uses the multiple self-attention layers to learn the historical stock movements while using the GCN to explore various relations among the stocks.
    \item \textbf{STHAN-SR}~\cite{sawhney2021stock} incorporates the Hawkes process to enhance the temporal evolution of stock movements, where a neural hypergraph is proposed to integrate the intra-domain knowledge.
\end{itemize}
In each baseline, we first produce the predicted results regarding return ratio and then yield a ranked list of stocks.

\begin{table*}[ht]
\caption{Performance comparison on NASDAQ and NYSE.}
	\renewcommand{\multirowsetup}
    {\centering}
\label{table:performance}
\centering
\footnotesize
\setlength{\tabcolsep}{1.5mm}{

\begin{tabular}{l|c|c|c|c|c|c|c|c|c|c|c|c}
	\toprule[1.0pt]
	\multicolumn{1}{l|}{\multirow{2}{*}{\textbf{Dataset}}}&
	\multicolumn{6}{c|}{\textbf{NASDAQ}}&\multicolumn{6}{c}{\textbf{NYSE}}\cr
	\cline{2-13} 
	\textbf{}  &\multicolumn{3}{c|}{\textbf{Weekly}}&\multicolumn{3}{c|}{\textbf{Monthly}}&\multicolumn{3}{c|}{\textbf{Weekly}}&\multicolumn{3}{c}{\textbf{Monthly}}\\
	\hline
	\multicolumn{1}{l|}{\textbf{Method}}& \multicolumn{1}{c|}{SR}  &  \multicolumn{1}{c|}{MRR}  & \multicolumn{1}{c|}{NDCG@5}  &     \multicolumn{1}{c|}{SR}   &   \multicolumn{1}{c|}{MRR}     &     \multicolumn{1}{c|}{NDCG@5} &  \multicolumn{1}{c|}{SR}       &    \multicolumn{1}{c|}{MRR}     &       \multicolumn{1}{c|}{NDCG@5}  &      \multicolumn{1}{c|}{SR}    &    \multicolumn{1}{c|}{MRR}     &     \multicolumn{1}{c}{NDCG@5}    \\
	\hline
ARIMA & 0.6484     &   0.0026    &   0.8090         &   0.6341     &     0.0076      & 0.7418&  0.1921  &   0.0032      &       0.8057     &   0.3313       &    0.0034      &  0.8236        \\
LSTM &  0.3135    &  0.0250       &  0.8301       & 0.2127         &  \underline{0.0342}     &  0.8148& 1.4760      &       0.0385  &    0.7735     &  0.4124        &    0.0212       &     0.7275      \\

GRU &   0.3937   &  0.0238       &  0.6919       &    0.2370      &  0.0249       & 0.7107 & -   &    0.0315     &   0.8774      &    0.4974      &  \underline{0.0426}  &   0.8005        \\

DA-RNN &  1.1010   &   0.0212     &    0.6667    &  0.2664        &   0.0341      & 0.8380 &    1.4318  &   0.0407      &     0.7435    &  \underline{1.5184}       &  0.0383       &   0.7232        \\

CNNpred &    0.6869   &    0.0168     &     \underline{0.8997}     &   0.7886      &    0.0175     & 0.8085 &   \underline{1.9209}   &  0.0241       &   \underline{0.8891}      &    0.7302        &    0.0329     &  0.6613        \\
StockNet &   0.8283   &   0.0325  &   0.6965  &   \underline{1.0999}   &  0.0128  &0.7307 & 0.4257   & 0.0434     &    0.8653     &-      &   0.0067  &0.8739       \\
\hline
LSTM+GCN & 0.9721    &   0.0259       &    0.8785      &     0.4929     &     0.0142     & \underline{0.8832} &    1.0316    &    0.0150      &      0.8462    &     1.4876      &    0.0046      &   \underline{0.8862}       \\

RSR &  0.5637    &    \textbf{0.0389}     & 0.8283    &      0.6876    &   0.0132      &0.8200 &  1.4715    &    \underline{0.0446}     &        0.7735     &     0.0453     &  0.0325       & 0.7849         \\

HATR &   1.4083    &    0.0210    &    0.7972     &    0.9403      &   0.0167     &0.8039  &0.5130      &  0.0424       &   0.8403      &  0.7699        &   0.0183      &    0.8549      \\

STHAN-SR &  \underline{1.4139}    &  0.0240       &   0.8012      &     0.9924     &    0.0190      & 0.8299 &  1.1901    &   0.0312      &  0.8857       &    \underline{0.4046}      &   0.0405      &   0.6496        \\
\hline
\hline
\textbf{StockODE} &   \textbf{1.6764}      &   \underline{0.0380}     & \textbf{0.9433}    &           \textbf{1.5840} &   \textbf{0.0343}     &\textbf{ 0.9432}&   \textbf{1.9843}   &    \textbf{0.0447}    &    \textbf{ 0.8892}     &   \textbf{1.6497}       &    \textbf{0.0442}     & \textbf{0.9420}         \\
	\bottomrule[1.0pt]
\end{tabular}}
  \begin{tablenotes}    
        \footnotesize               
        \item[1]   Herein,  `-' means that the indicator is negative. For each method, we report average results over five runs.    
      \end{tablenotes}            
    
\end{table*}

\subsubsection{Metrics.}
Following~\cite{Feng2019TemporalRR,sawhney2021stock}, we evaluate all methods with three common metrics for stock investment decision: \textit{Sharpe Ratio (SR)}, \textit{Mean Reciprocal Rank (MRR)} and \textit{NDCG@K}.

SR is to measure the return of a portfolio against its risk. The ratio is the average earned return per unit of volatility or total risk over the risk-free rate, which can be calculated as follow:
\begin{equation}
    \label{SR}
    \text { SR }=\frac{R_{p}-R_{f}}{\sigma_{p}},
\end{equation}
where $R_{p}$ represents return of portfolio investment, $R_{f}$ represents risk-free rate, and $\sigma_{p}$ represents standard deviation of the portfolio excess returns. We use this classical ratio to show the return level of those stocks that are ranked and recommended by our model after balancing the risk. In our experiments, we attempt to analyze the return effect of selected results by showing the average return of the selected top five stocks' SR.

As we aim to select the most profitable stocks for investment, several error-based metrics such as RMSE (Root Mean Square Error) and MAE (Mean Absolute Error) are not suitable for evaluating the ranking performance. We follow previous studies~\cite{hsu2021fingat,Feng2019TemporalRR} and choose MRR as one of the significant metrics, which evaluates the predicted rank of the top-1 return ratio stock in the ground truth. To this end, MRR is defined as:
\begin{equation}
    \label{MRR}
    \mathrm{MRR}=\frac{1}{\mathcal{T}} \sum_{\tau=1}^{\mathcal{T}} \frac{1}{\operatorname{rank}_{\tau}},
\end{equation}
where $\operatorname{rank}_{\tau}$ returns the real position of the predicted highest-ranked stock in the ground truth on the $\tau$-th testing day and $\mathcal{T}$ is the number of testing days.

Additionally, to display the effectiveness of the portfolio, we also choose NDCG@K as the last metric for model evaluation. NDCG@K is a widely used metric to measure ranking quality. Following~\cite{sawhney2021stock}, we report the evaluation results of NDCG@5 in this paper.

\subsubsection{Implement Details}
We implemented our StockODE and baselines in Python where the deep learning-based methods are built upon the PyTorch library, accelerated by the NVIDIA RTX 3090 GPU 24G. Specifically, the dimensionality of hidden state in Movement Trend Correlation is 64. The dimensionality of hidden state in NRODE is 64. The output space in our hierarchical hypergraph is 32. We use the popular Adam optimizer for StockODE training, where the initial learning rate is 0.001.
Our evaluation function in ODE solvers is three-layer MLPs with 64 hidden units in each layer. We empirically set $\beta=0.1$. 

\subsection{Comparison and Analysis}
\subsubsection{Overall Performance}
Table~\ref{table:performance} presents the stock ranking performance of different methods including ours. Specifically, we respectively report the results conducted on weekly-level and monthly-level stock movements. We note that the best gain is shown in bold, and the second best is shown as underlined. 

\textbf{Baseline Performance.} Among the non-graph-based baselines, we first find that ARIMA as a simple time series forecasting model achieves good results in terms of NDCG@5. Especially, ARIMA even performs better than RNN-based (e.g., LSTM and DA-RNN) methods, which demonstrates that there exist significant temporal dependencies behind the historical stock transactions. However, we observe that the performance of ARIMA in different contexts has obvious fluctuations regarding SR and MRR. We also find that LSTM and GRU have similar observations when tackling the different datasets or the different levels of historical stock movements. We consider the plausible reason is that stock movements are usually affected by multiple factors besides inherent temporal dependencies, as revealed by the Efficient Market Hypothesis (EMH). Compared to GRU, DA-RNN with the attention mechanism performs better regarding SR, which indicates that learning the potential hierarchical interactions behind the temporal dependencies does help make a promising trade-off between the risk and return. For StockNet, it employs the neural variational inference to understand the stock movements by addressing the intractable posterior inference, which archives more stable results than other RNN-based methods. Among the graph-based baselines, these methods obtain more robust and higher gains than solely RNN-based methods, which demonstrates that incorporating the high-order dependencies among stocks does facilitate the recommendation of more profitable stocks. For LSTM+GCN and RSR, they devise the graph convolutional neural network-based models to incorporate the corporation relationships, yielding more stable performance when tackling different temporal levels of historical stock transactions. For instance, RSR achieves the best results in terms of MRR when the lookup window is at the weekly level. STHAN-SR is a recent attention-based model that uses the vanilla neural hypergraph network to consider the intra-domain knowledge behind the corporation relationships, which achieves the best SR results against the other graph-based methods. However, it does not bring more effective ranking quality as it performs poorly on MRR and NDCG@5. 

\textbf{Our solution.} We can observe that our proposed StockODE significantly outperforms all the baselines over both two datasets except for weekly-level movements in NASDAQ regarding MRR, which demonstrates that StockODE is a more effective method to tackle dependencies behind the stock movements as well as uncover the complex interactions among the stocks. Regarding the longitudinal comparison, SR shows the return level of the selected stocks and comprehensively considers the level of risk. The higher SR, the better-selected stocks are obtained. Compared with other baselines, our model obtained is well behaved with SR. MRR and NDCG@5 show the difference between the selected stocks and the real stock ranking. They can provide a basis for the selection of stocks to a certain extent. Overall, the prediction difference between the stocks recommended by the model and the best ranking of real stock returns is the smallest, so the recommended ratios are the highest. We will conduct more empirical investigations on each of the design components of StockODE in the following to evaluate their distinctive contributions.

\begin{figure}[ht]
    \centering
\subfigure[SR.]{\includegraphics[width=0.495\textwidth]{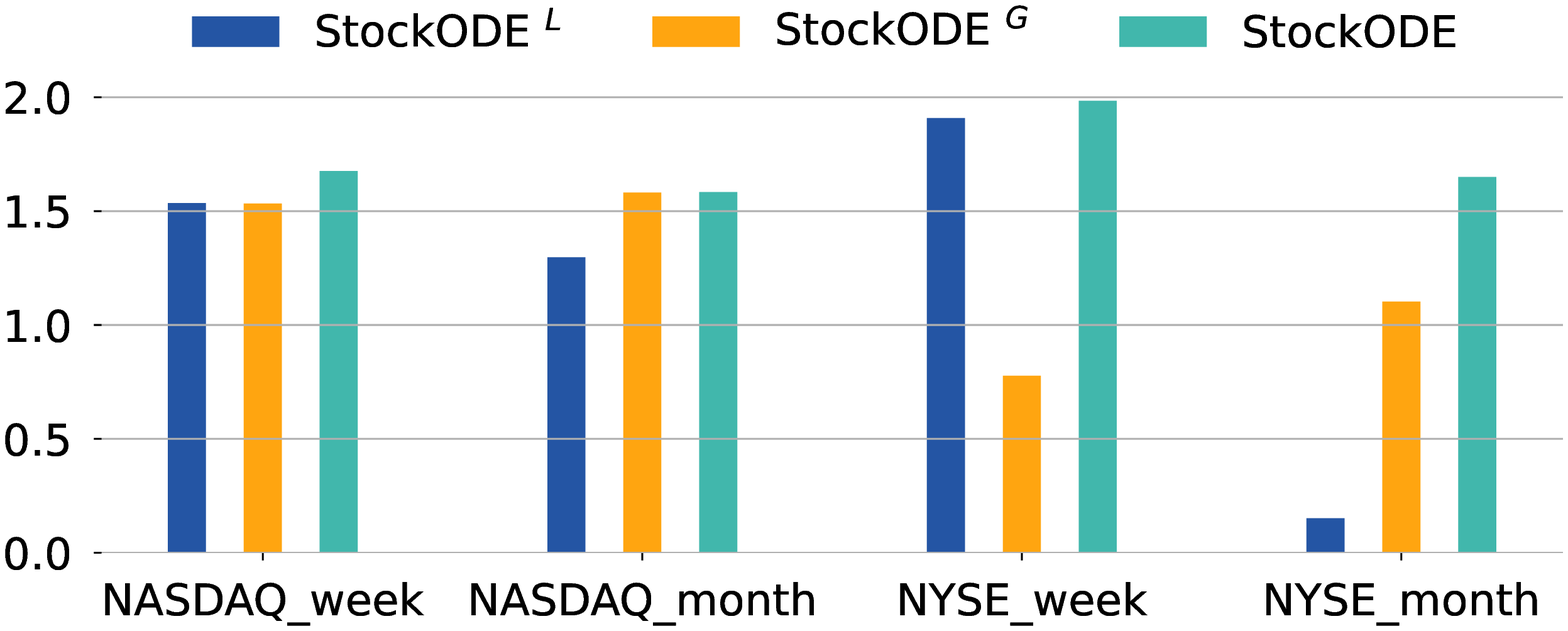}
\label{SR-res}}
\subfigure[MRR.]{\includegraphics[width=0.495\textwidth]{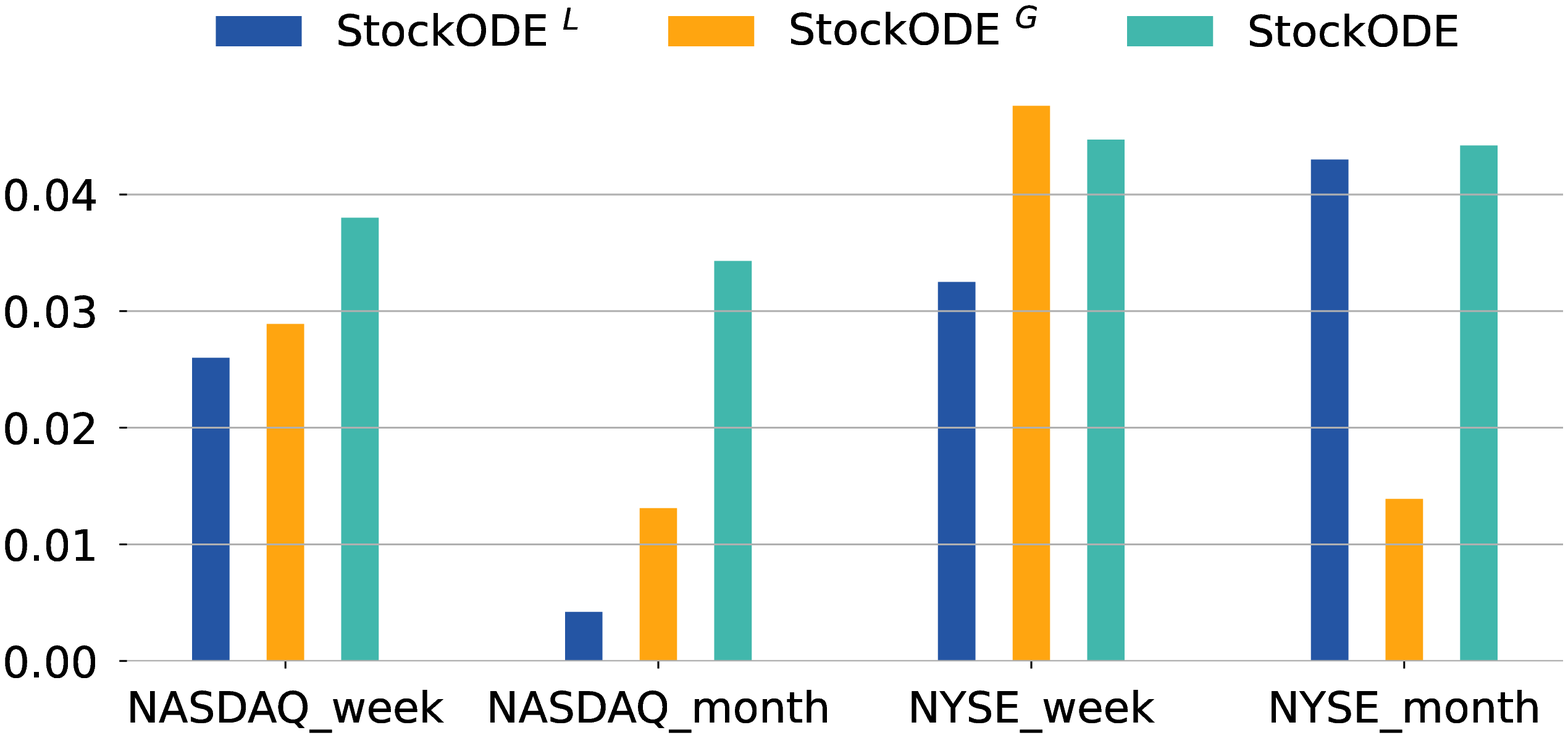}
\label{MRR-res}
}
\subfigure[NDCG@5.]{\includegraphics[width=0.495\textwidth]{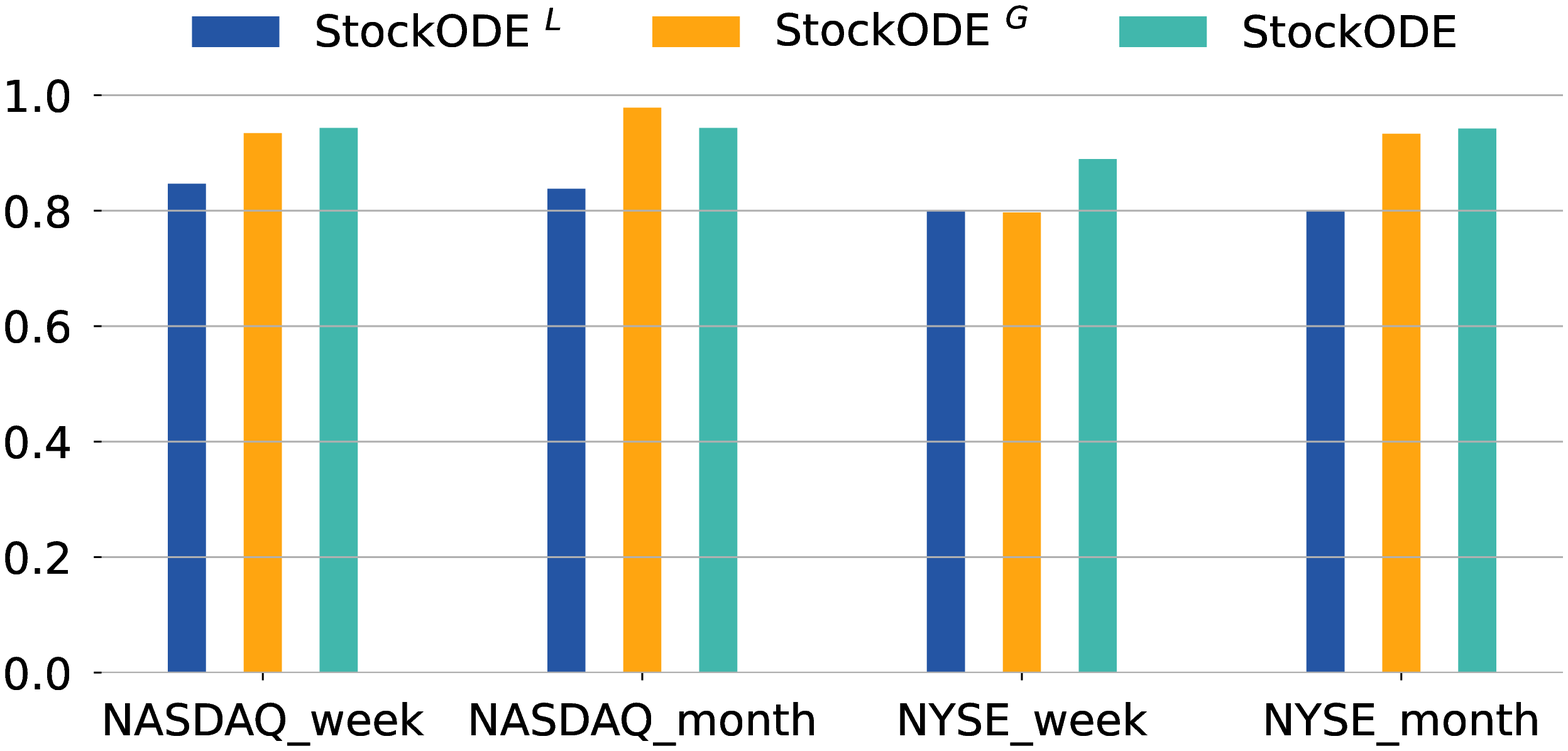}
\label{NGCG-res}
}
\caption{ODE comparison.}
\label{fig-odes}
\end{figure}

\textbf{ODE Comparison.} We also select the most recent autoregressive models with ODE as the variants of our StockODE. Specifically, we adopt the Latent ODE~\cite{chen2018neural} for stock movement learning where the ODE is only used in the reconstruction part. And we define it as $\textbf{StockODE}^L$. In addition, we use the ODE-RNN~\cite{rubanova2019latent} as our encoder part, and we call this model as $\textbf{StockODE}^G$ where we select the GRU as the recursive cell. As shown in Fig.~\ref{fig-odes}, It is clear that most evaluation results obtained from our StockODE achieve the best performance compared to traditional ODEs, which indicates that our devised NRODE is a reasonable and competitive time-dependent learning module.
\begin{figure}[ht]
    \centering
    \includegraphics[width=0.5\textwidth]{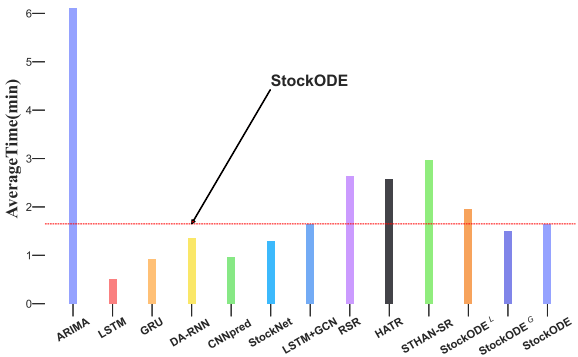}
    \caption{Training time comparison.}
    \label{fig:runtime}
\end{figure}

\textbf{Model Efficiency.} Finally, we visualize the training time of the baselines as well as our StockODE. As shown in Fig.~\ref{fig:runtime}, the blue one represents the training time after reaching optimal performance, and the orange one represents the average time of 50 training epochs. We can observe that StockODE obtains competitive performance in terms of training time.

\begin{table}[ht]
\centering
\caption{Experimental results on the NASDAQ dataset at the weekly level.}
\label{table:ablation}
\footnotesize
\begin{tabular}{lccc}
\toprule
\multicolumn{1}{l|}{Method}& \multicolumn{1}{c|}{SR}  &  \multicolumn{1}{c|}{MRR}  & NDCG@5   \\
\hline
\multicolumn{1}{l|}{GRU}  & \multicolumn{1}{c|}{0.3937}     &     \multicolumn{1}{c|}{0.0238}      & 0.6919 \\
\hline
\multicolumn{1}{l|}{StockODE-B}  & \multicolumn{1}{c|}{1.2228}     &     \multicolumn{1}{c|}{0.0346}      & 0.8665 \\
\hline
\multicolumn{1}{l|}{StockODE-I}  & \multicolumn{1}{c|}{0.5004}& \multicolumn{1}{c|}{0.0334}         &0.9299 \\
\hline
\multicolumn{1}{l|}{StockODE-H} & \multicolumn{1}{c|}{1.4508}    &    \multicolumn{1}{c|}{0.0320}       &  0.9432 \\
\hline
\multicolumn{1}{l|}{StockODE-A}
&\multicolumn{1}{c|}{1.1182}&
\multicolumn{1}{c|}{0.0317}&
0.8812
\\
\hline
\hline
\multicolumn{1}{l|}{StockODE}  & \multicolumn{1}{c|}{1.6764}     &     \multicolumn{1}{c|}{0.0380}      & 0.9433  \\
\bottomrule
\end{tabular}
\end{table}
\begin{figure}[ht]
    \centering
    \includegraphics[width=0.485\textwidth]{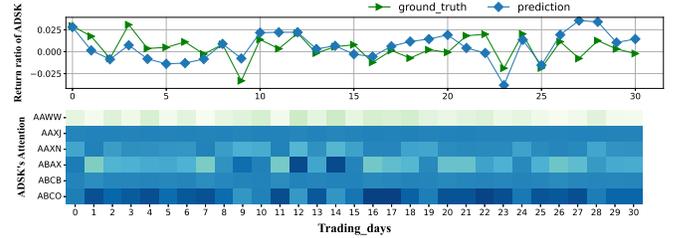}
    \caption{The visualization of implicit correlation in StockODE.}
    \label{fig:mtc}
\end{figure}
\subsubsection{Ablation Analysis}
Now we further conduct ablation analysis to investigate the contribution of each significant component in our StockODE. The variants of StockODE are as follows:
\begin{itemize}[leftmargin=*]
    \item \textbf{StockODE-B}: It uses our Movement Trend Correlation module for capturing the correlations among different stocks regarding the time-evolving movements, and then the standard GRU is employed for the stock movement modeling.
    \item \textbf{StockODE-I}: This variant does not consider the interactive facts in StockODE. 
    \item \textbf{StockODE-H}: It does not consider the Inter-domain knowledge part in hypergraph learning.
    \item \textbf{StockODE-A}: It replaces our NRODE in StockODE with the standard GRU for stock movement pattern learning. 
\end{itemize}
As shown in Table~\ref{table:ablation}, we also report experimental results obtained from a simple GRU~\cite{shen2018deep}. Generally, we can observe that the experimental results regarding each variant are significantly distinguishable, which demonstrates that each devised component does help facilitate task performance. In detail, StockODE-B outperforms GRU, which demonstrates that correlating the movement trends among different stocks is a positive signal to promote the selection of profitable stocks. StockODE-I performs worse than StockODE especially on SR, which indicates that incorporating the higher-order stock relations from multiple domains is capable of providing higher returns as well as relieving the investment risks. For StockODE-H, it performs worse than our StockODE, which demonstrates that inter-domain knowledge is also a key signal for promoting stock ranking. The plausible reason is such a piece of knowledge learned by our hypergraph network is capable of exploiting the implicit stock correlations. Finally, StockODE outperforming StockODE-A reveals that our proposed NRODE is a highly predominant module that enables dynamically the temporal dependencies from the historical stock movements.

\begin{figure}[t]
    \centering
    \includegraphics[width=0.47\textwidth]{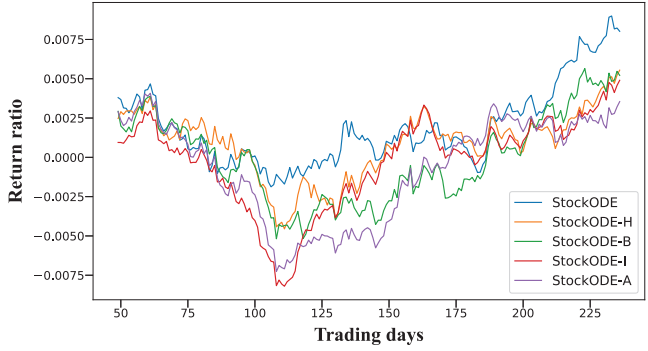}
    \caption{The comparison of the average returns of the top five stocks.}
    \label{fig:continuity}
\end{figure}

As shown in Fig~\ref{fig:mtc}, we take stock \textit{AAWW} as an example to show the results of our Implicit Correlation Aggregation in StockODE, as well as the prediction results compared to the ground truth. As the top of Fig~\ref{fig:mtc} presents, our StockODE is able to predict return ratios that closely match the ground truth over the long term, especially for the evolutionary trends. During inference, we visualize part of the attention scores of \textit{AAWW} regarding other stocks. The darker the color, the higher the attention score. We can find that their scores evolve over time, suggesting that considering time-varying correlations can help us reveal the strength of the relationship between different stocks in the time domain.

In the end, we use each of the above models including our StockODE to predict a ranked list and then use each list to evaluate investment returns.
Specifically, we use the real stock prices of the five selected stocks to calculate the daily return ratio, and then judge the profit margin brought by the selection results for investors. Fig.~\ref{fig:continuity} shows the average return ratios of the top five stocks over the 50 testing days. We can find that the selected results from StockODE bring us higher profits on most of the trading days than other methods.

\subsubsection{Continuity Analysis}

\begin{figure}[ht]
    \centering
    \includegraphics[width=0.48\textwidth]{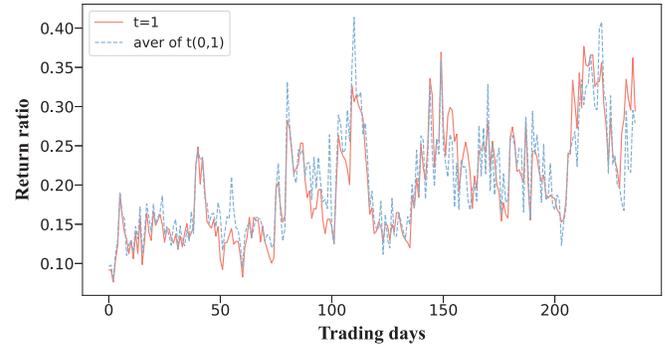} 
    \caption{The evolution of long-term daily average returns.}
    \label{fig:long-return}
\end{figure}

\begin{figure}[ht]
    \centering
\subfigure[short-term.]{    \includegraphics[width=0.17\textwidth]{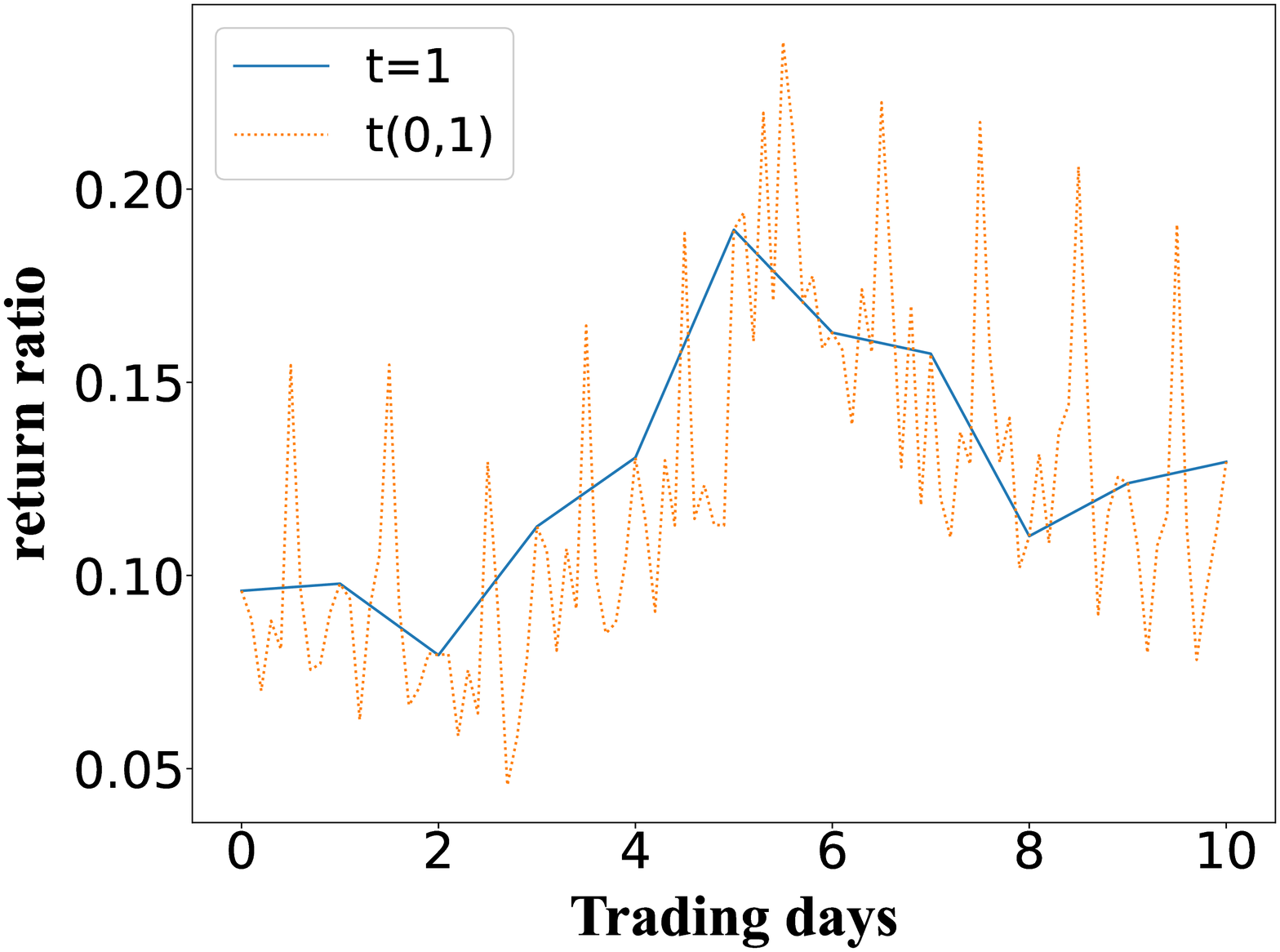}
}
\hspace{-0.7cm}
\subfigure[middle-term.]{    \includegraphics[width=0.17\textwidth]{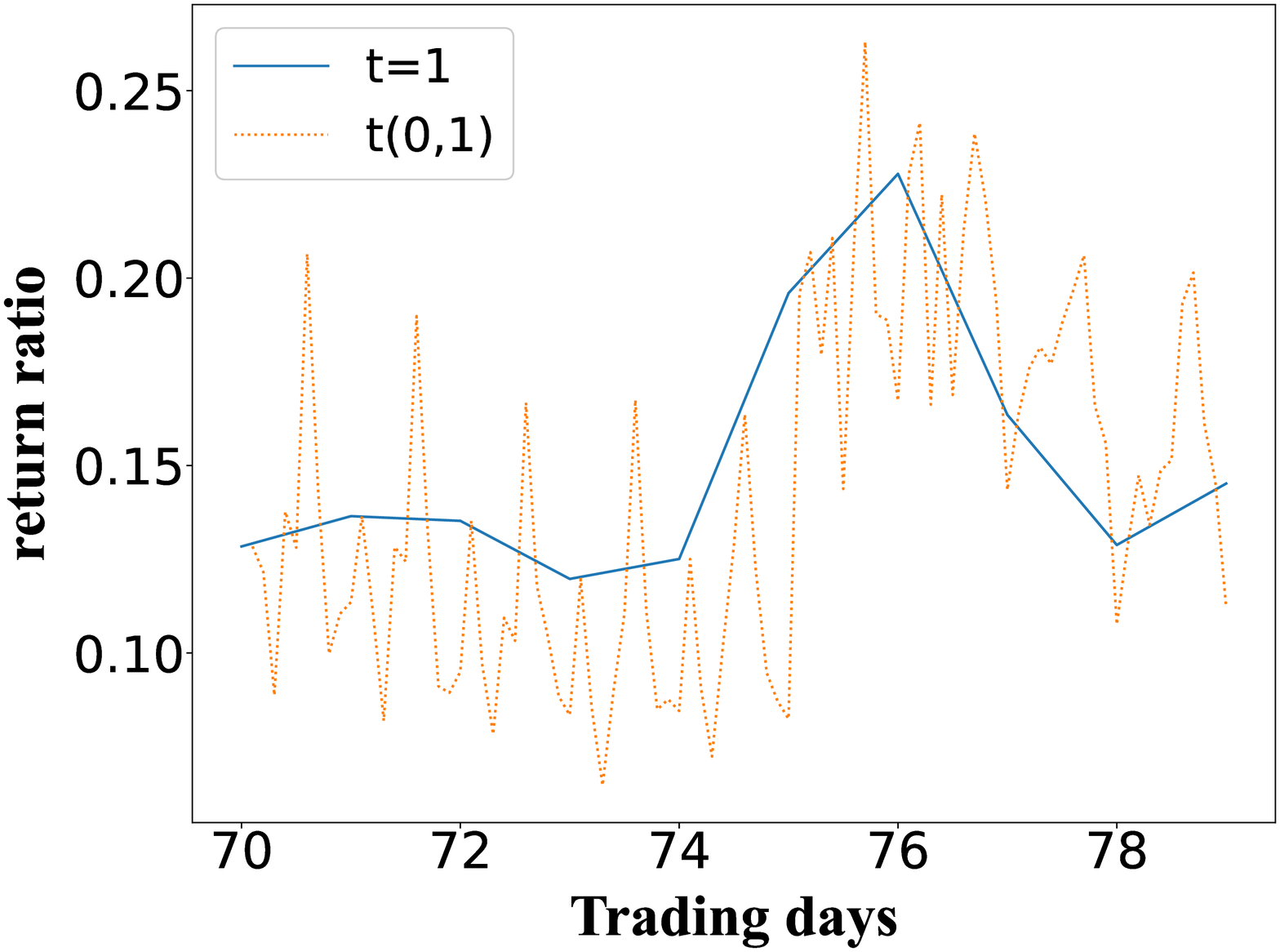}
}
\hspace{-0.7cm}
\subfigure[long-term.]{    \includegraphics[width=0.17\textwidth]{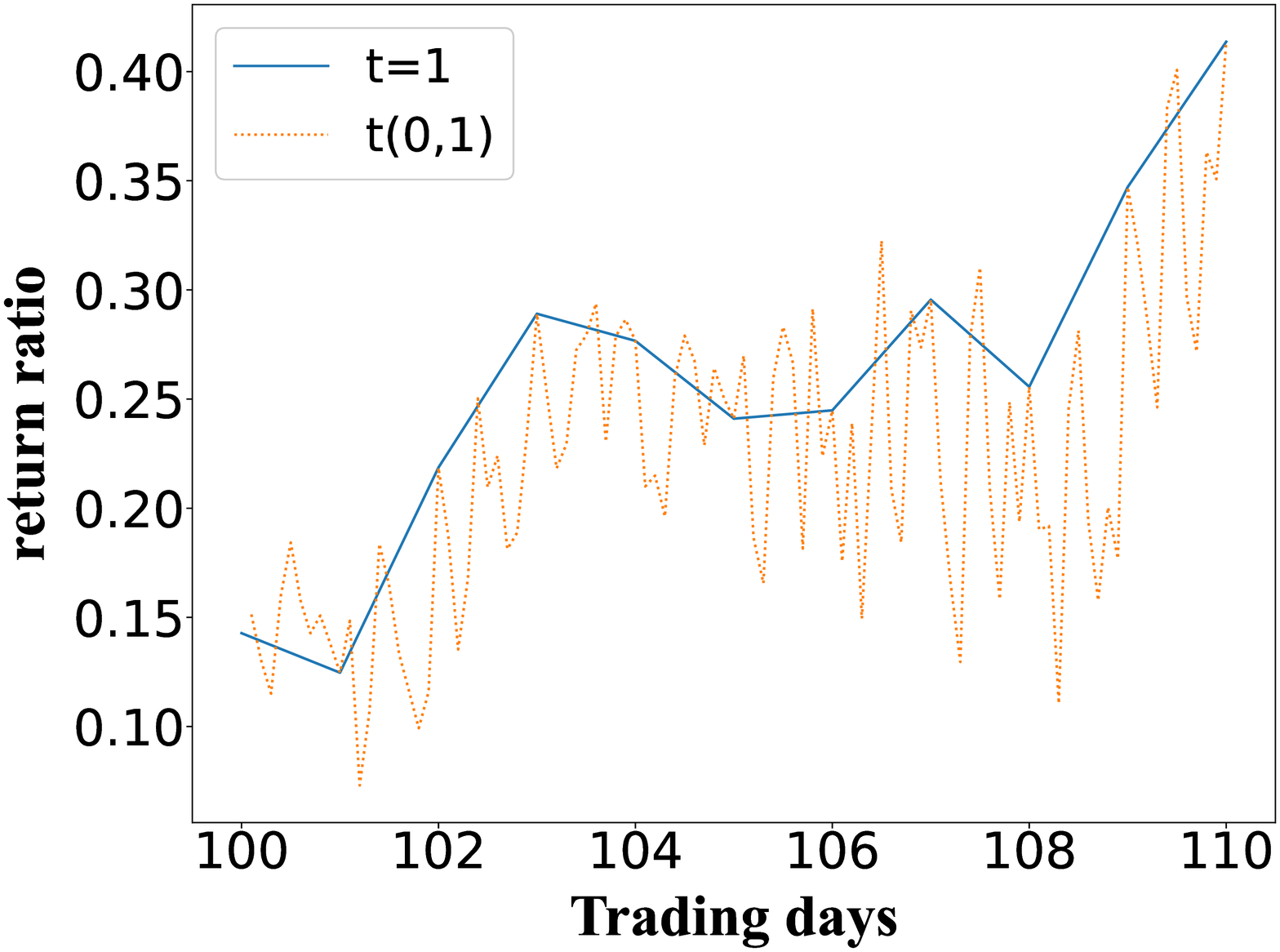}
}
    \caption{
The evolution of average returns over different time horizons.
}
\label{fig:avg_time}   
\end{figure}

Now we turn to make continuity analysis to demonstrate the advantage of ODE regarding the stock market fluctuations. 
As one of the inherent advantages, StockODE follows the paradigm of continuous dynamics systems, which can generate flexible but extensive results with varying observation time intervals. We visualize averaged return ratio of the top-five stocks of NASDAQ selected by our StockODE. As shown in Fig.~\ref{fig:long-return}, the red curve is the ground-truth results based on selected stocks. And we use the StockODE with 10 different time steps, ranging from (0.0,1.0), to generate a more fine-grained curve to mimic the changes of these stocks, we can find the blue curve significantly approximates the ground truth across long-term testing days. As we mentioned in Sec.~\ref{Introduction}, the stock prices could fluctuate rapidly within minutes or hours, traditional RNN-based model can only infer the results with a discrete fixed time interval (i.e., daily observation.). In contrast, our StockODE is capable of generating observations at any flexible time. That is to say, StockODE can generate more fine-grained stock movements and incorporate immediate volatility and uncertainty in movements to help investors mitigate investment risks. As shown in Fig.~\ref{fig:avg_time}, we visualize the predicted evolution of average returns with a fine-grained time step, i.e., 0.1. We find that the results in different time horizons have obvious disturbances within each test trading day, suggesting that our StockODE is capable of simulating the uncertainty of stock movements.

\begin{figure}[ht]
    \centering
    \includegraphics[width=0.5\textwidth]{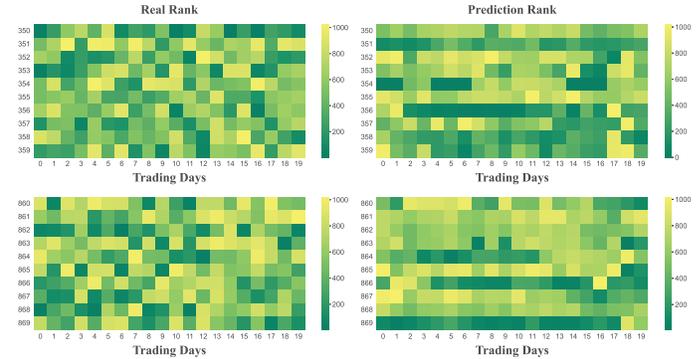}
    \caption{
The heat map of the ranking results of the selected 20 stocks over the first 20 testing days. Specifically, the left figure is the best-ranking position of each stock according to each day's stock price, the right figure presents the predicted ranking results.}
\label{fig:heat}
\end{figure}

Since most investors may pay attention to the investment returns brought by short-term fluctuations in the stock market and ignore the long-term returns of stock investments, resulting in potential investment risks. According to the left part of Fig.~\ref{fig:heat}, we generate the daily best rankings by calculating the return ratio of the stocks for each test trading day, we can find the ranking score of each selected stock shifts frequently due to the fluctuations of stock prices, which could confuse investors when picking potentially profitable stocks to invest in. In contrast, the results of the right part of Fig.~\ref{fig:heat} derived from our StockODE demonstrate that the predicted ranking score of each stock is significantly persistent. The reason is that StockODE coupling with ODE can consider stock trend fluctuations over a long period of time to relieve the risk of unreasonable decision-making and investment losses.

\begin{figure}[ht]
    \centering
    \subfigure[The metric results.]{\includegraphics[width=0.225\textwidth]{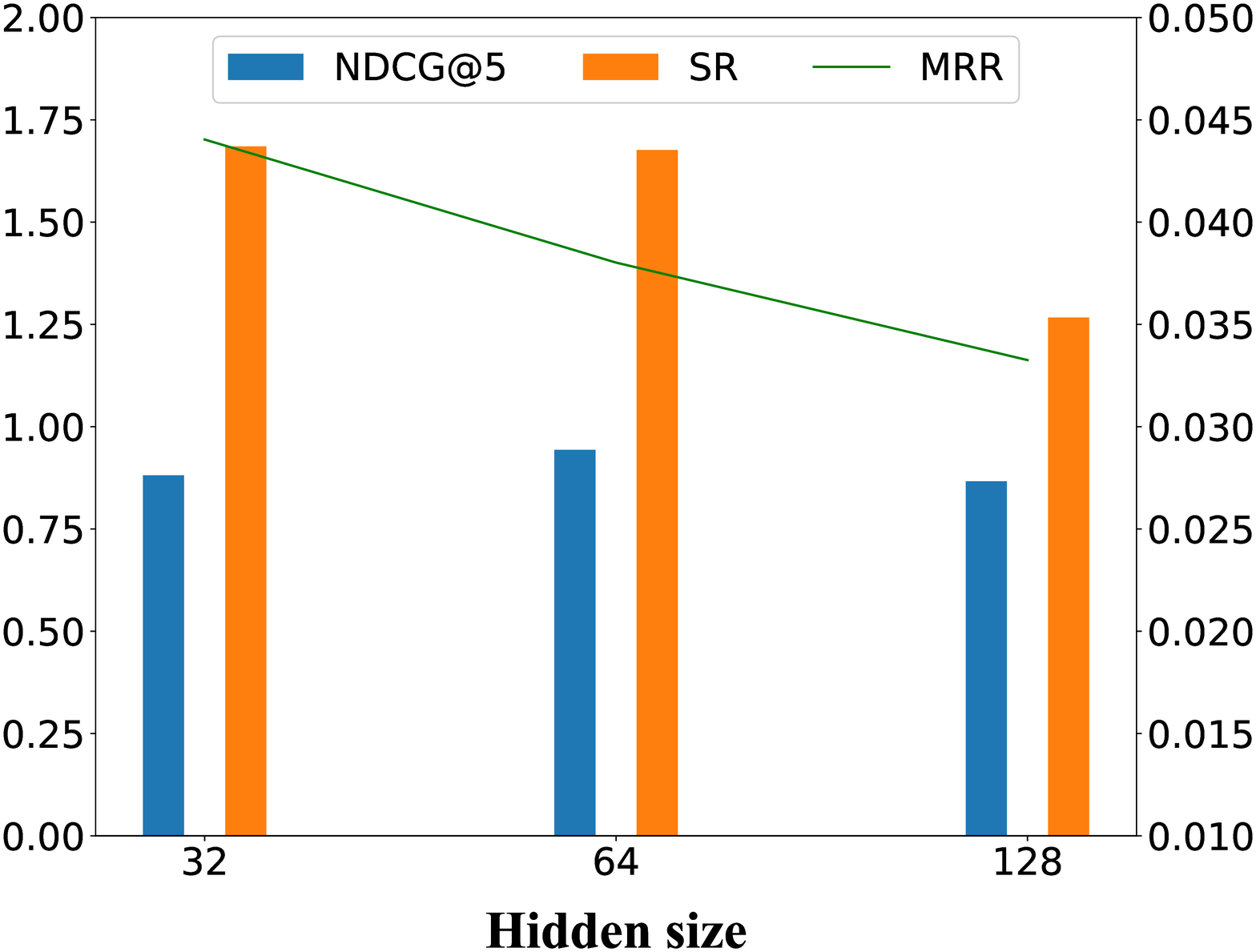}
    \label{fig:hidden-size}}
    \subfigure[The movements of returns.]{\includegraphics[width=0.225\textwidth]{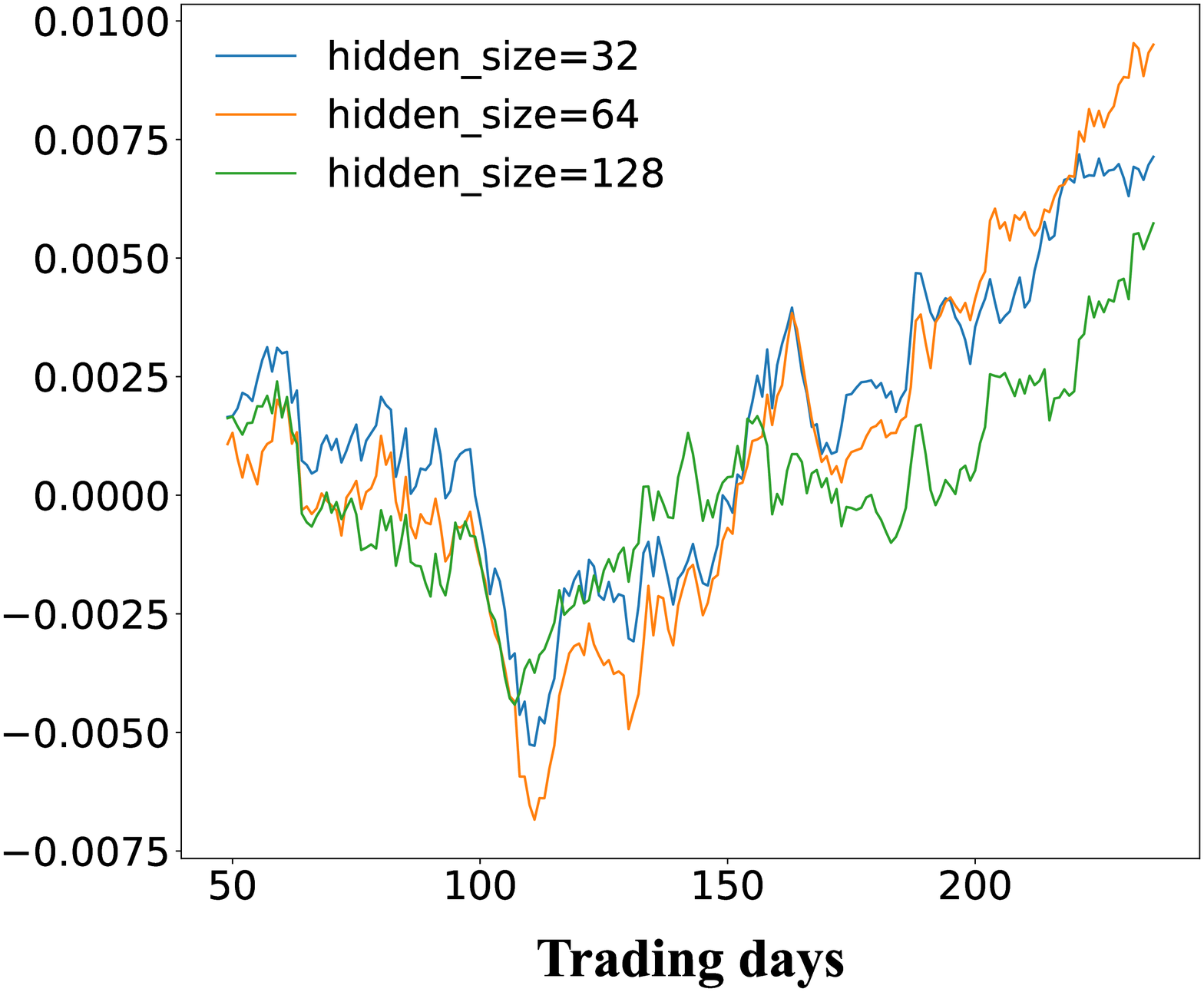}
    \label{fig:hidden-size_a}}
    \subfigure[The metric results.]{\includegraphics[width=0.225\textwidth]{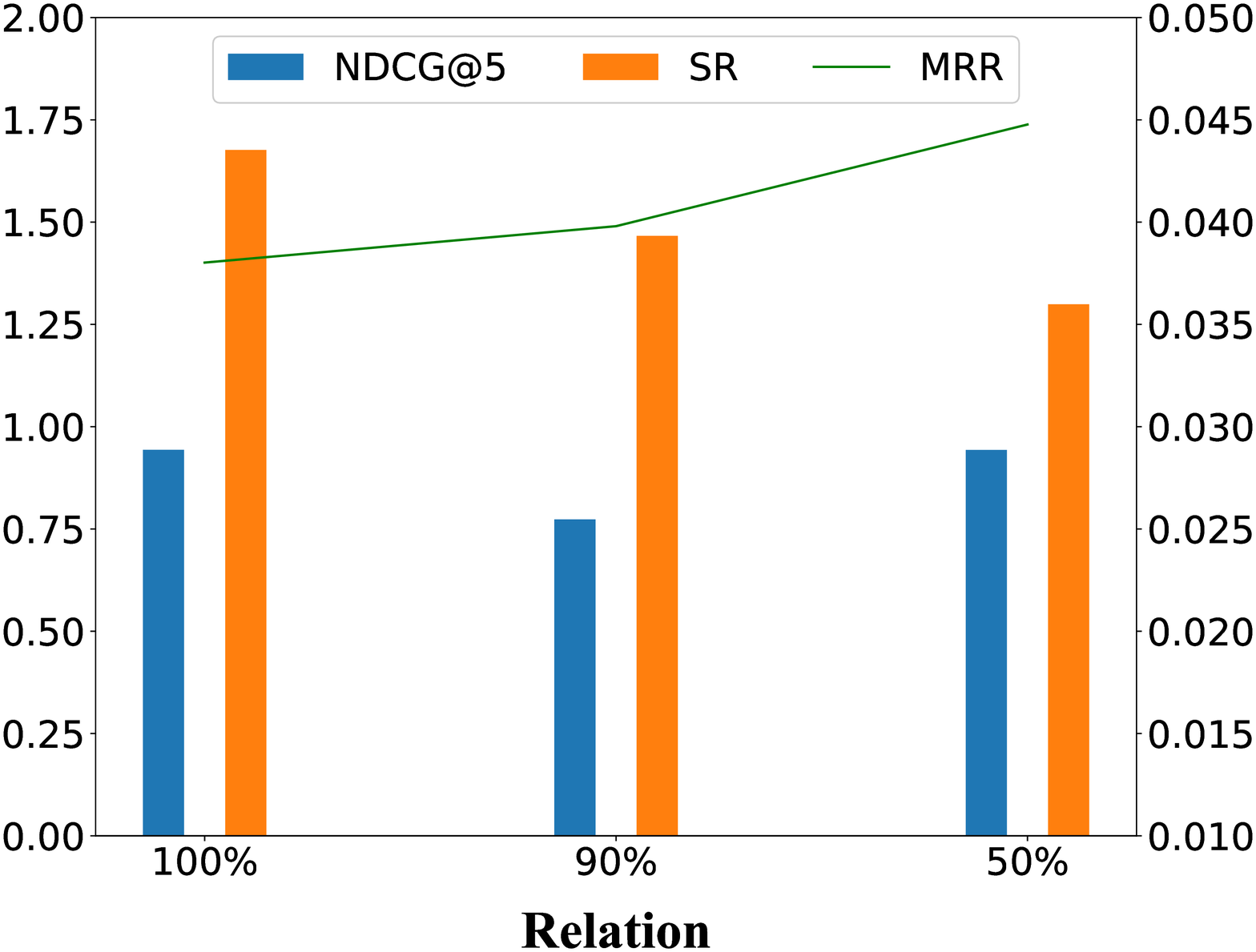}
    \label{fig:relation}}
    \subfigure[The movements of returns.]{\includegraphics[width=0.225\textwidth]{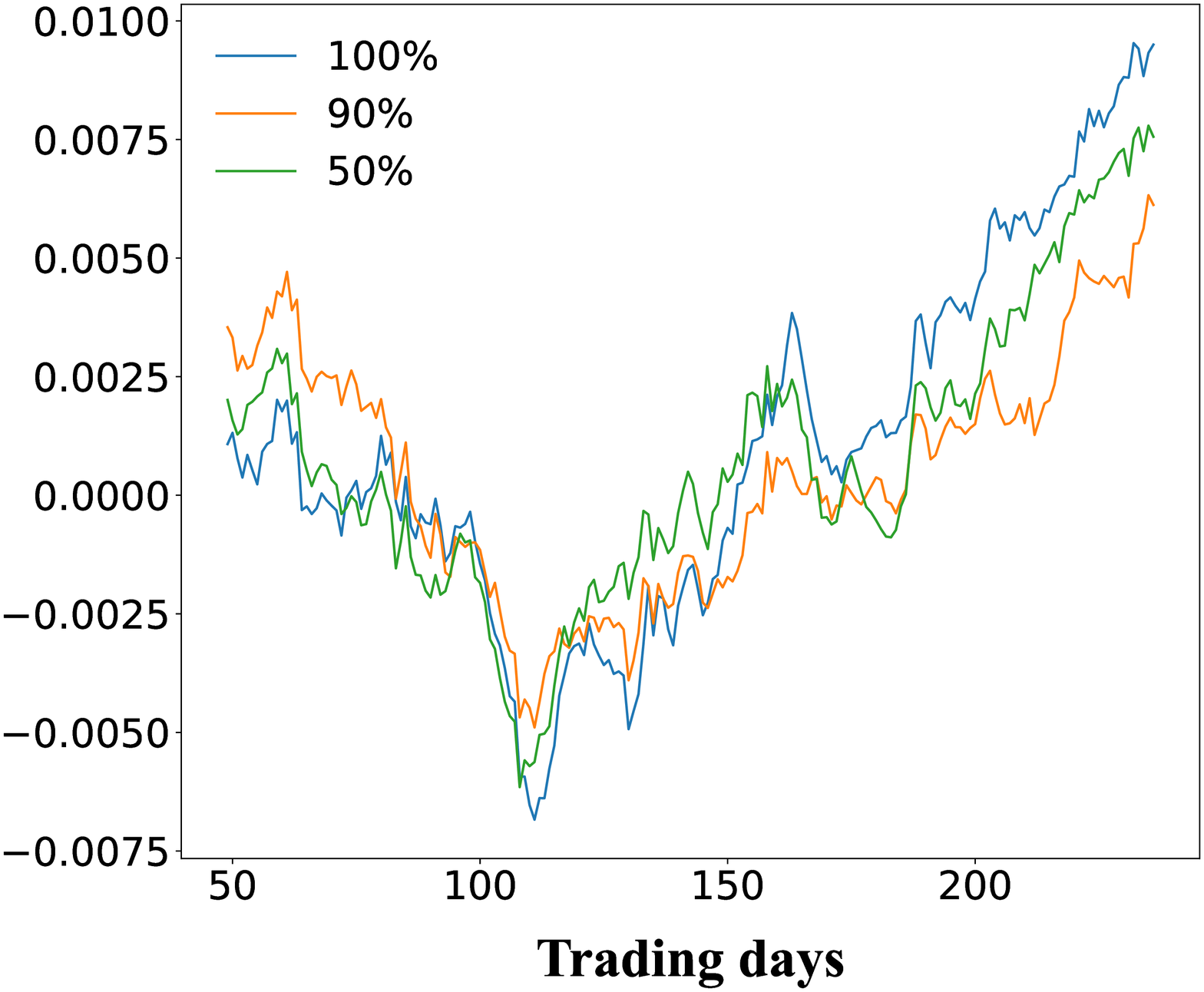}
    \label{fig:relation_a}}
    
    \subfigure[The metric results.]{\includegraphics[width=0.225\textwidth]{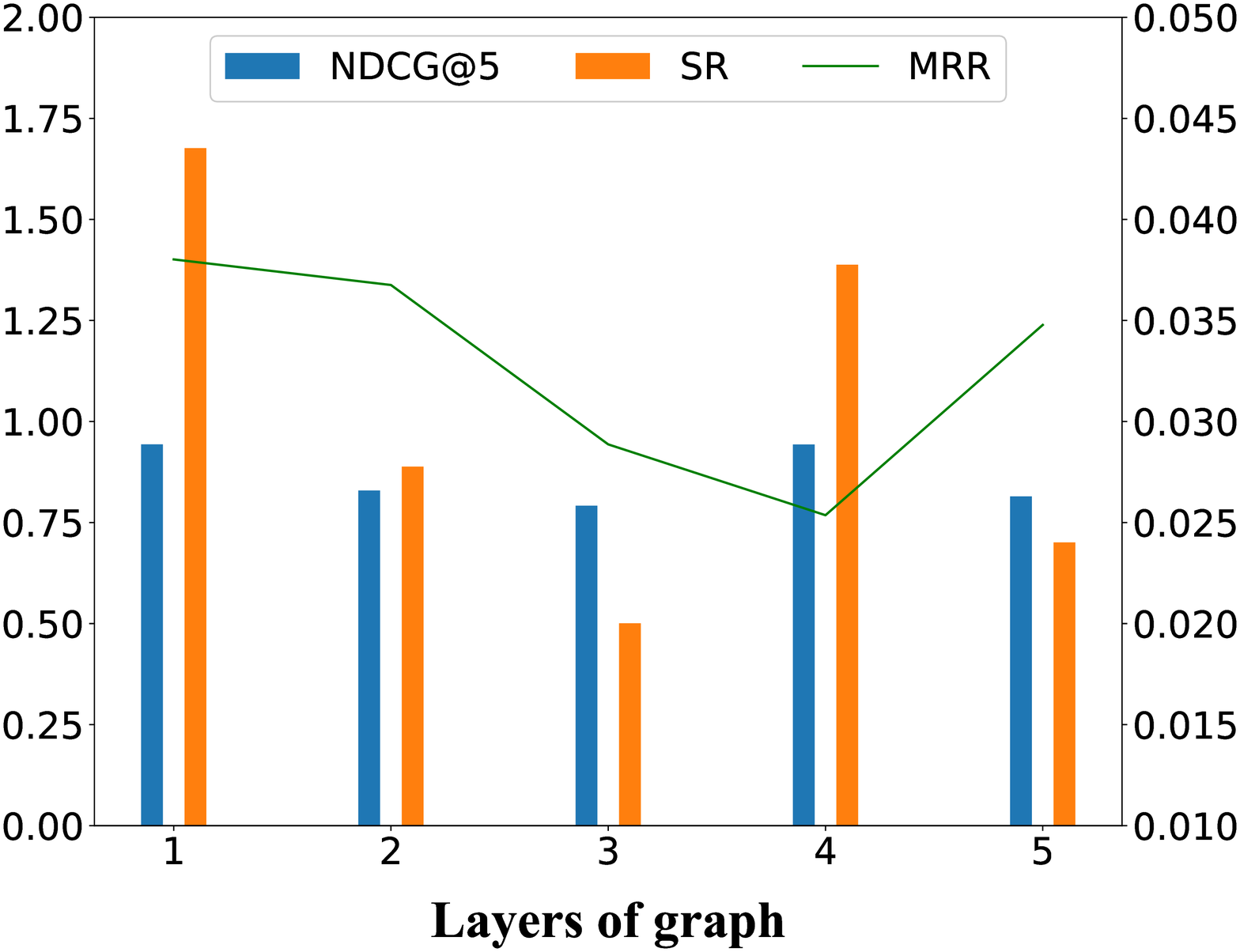}
    \label{fig:Layersofgraph}}
    \subfigure[The movements of returns.]{\includegraphics[width=0.225\textwidth]{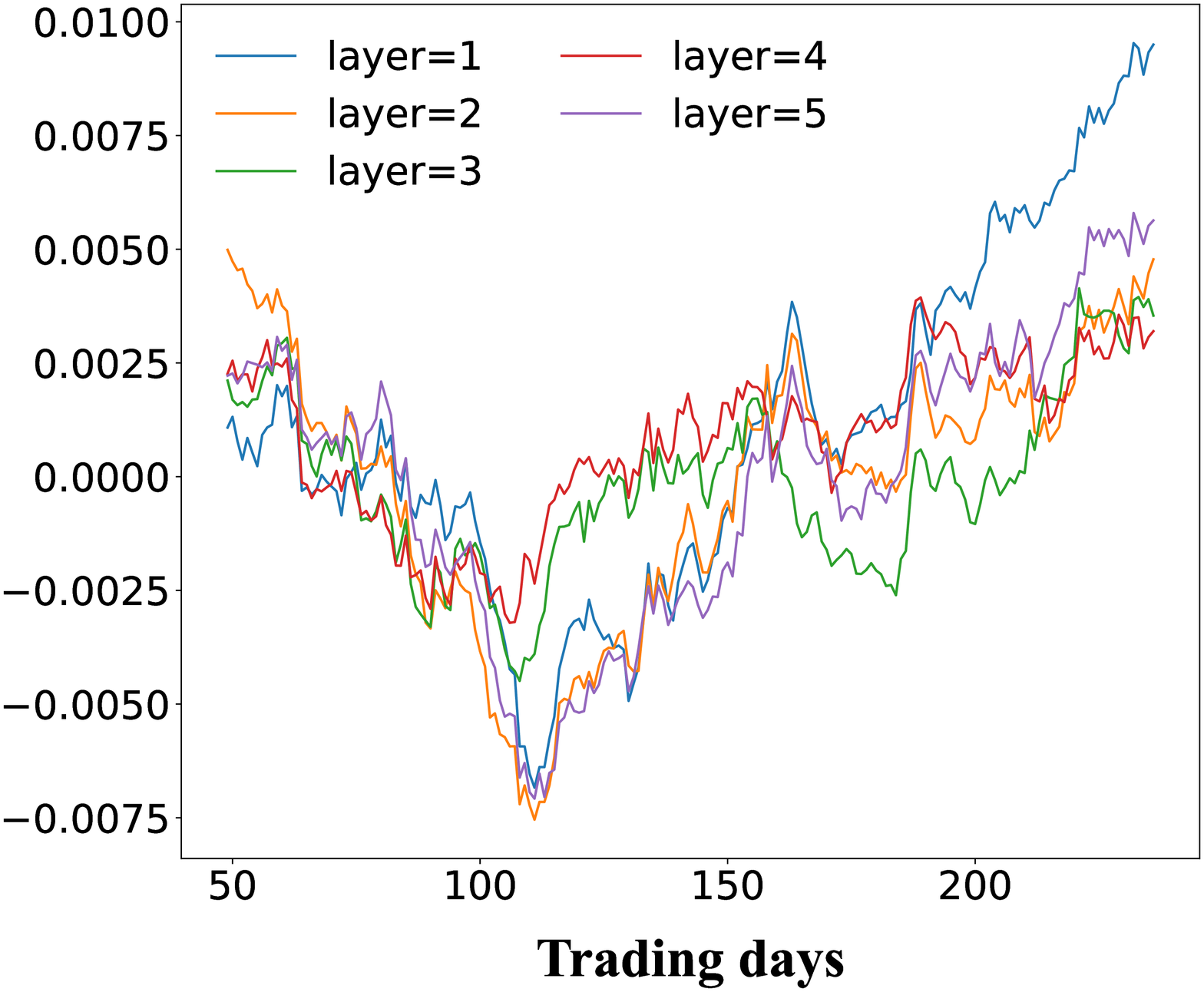}
    \label{fig:Layersofgraph_a}}
    \subfigure[The metric results.]{\includegraphics[width=0.225\textwidth]{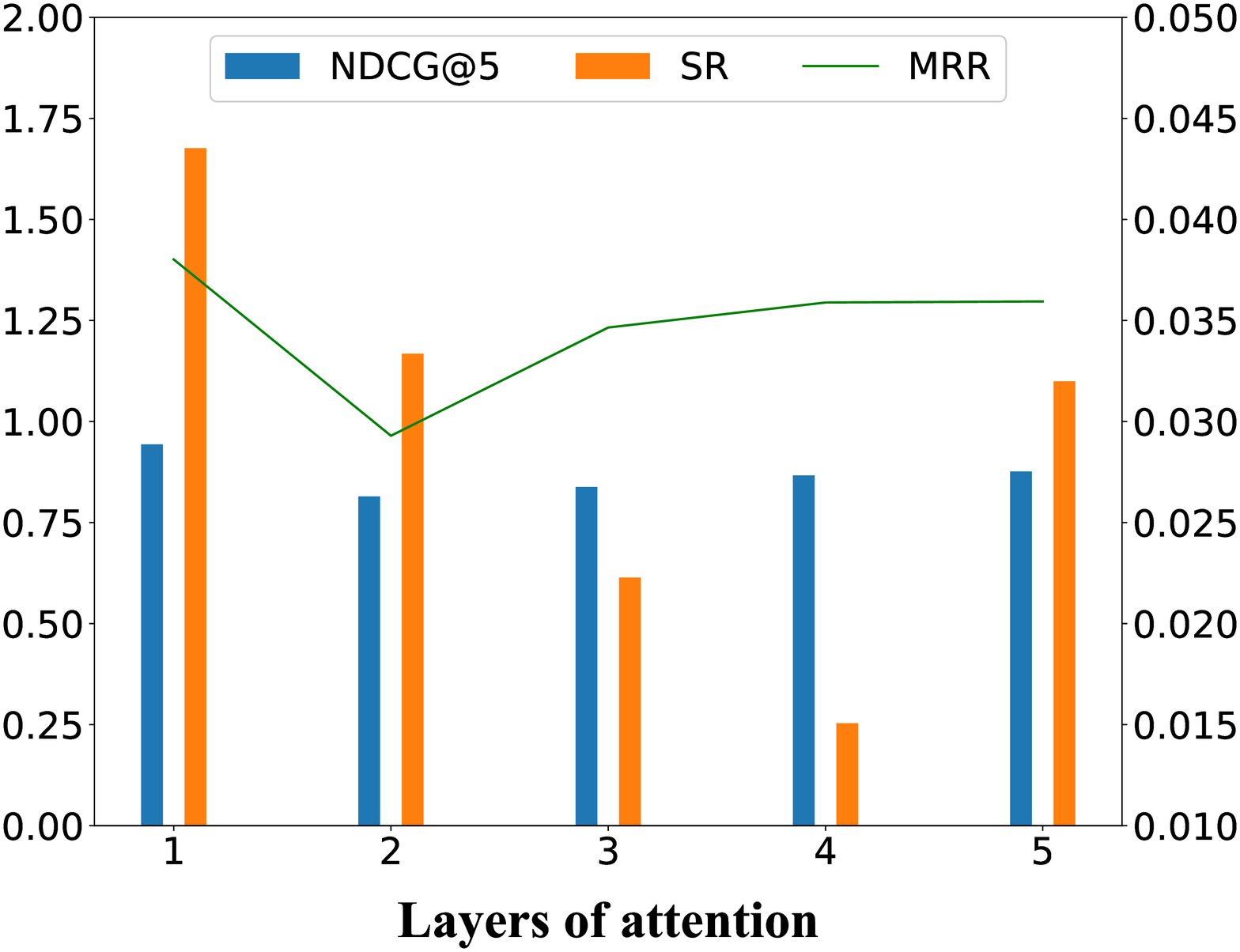}
    \label{fig:att}}
     \subfigure[The movements of returns.]{\includegraphics[width=0.225\textwidth]{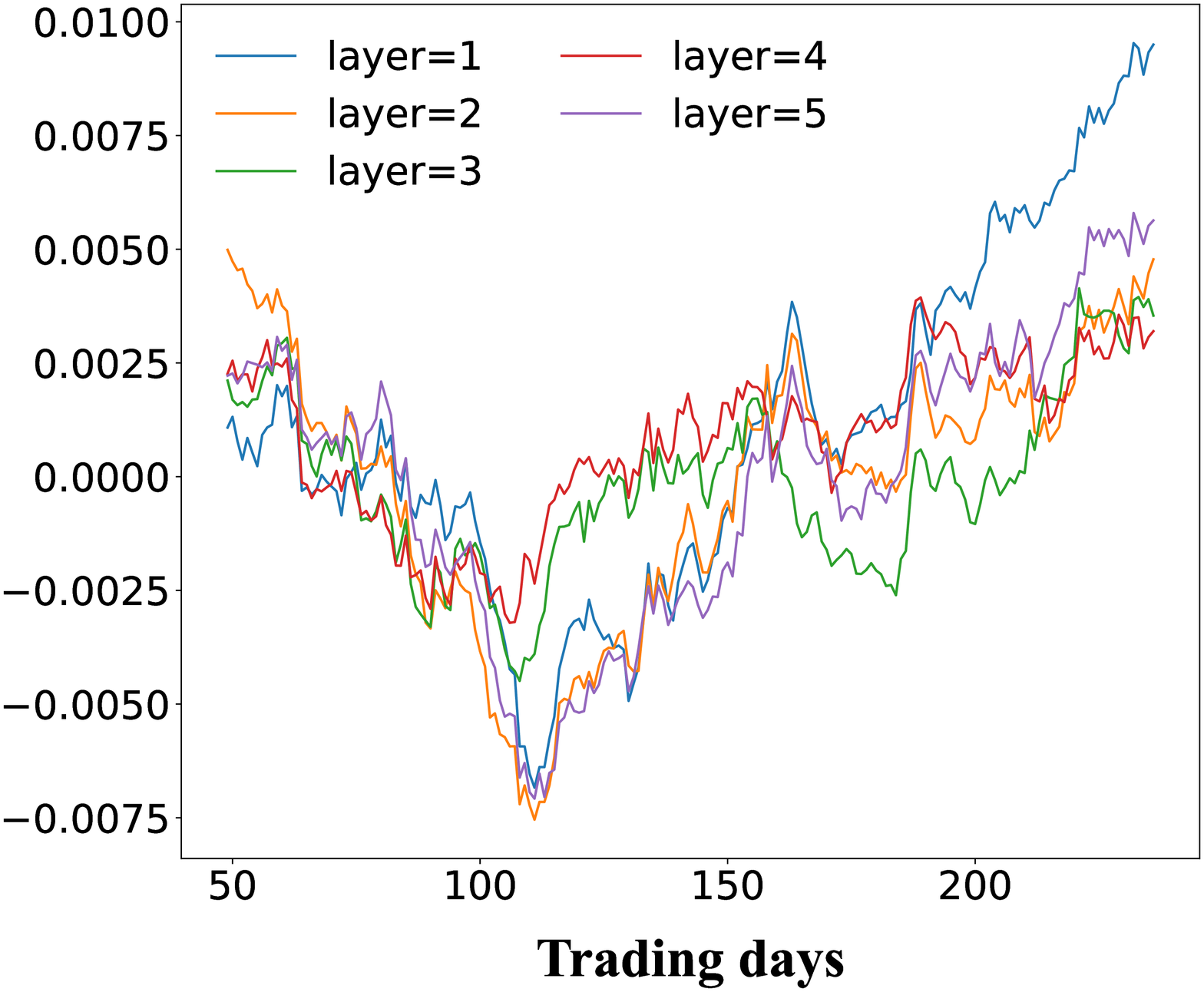}
    \label{fig:att_a}}
    \caption{
    Sensitivity Analysis on NASDAQ at the weekly level.
}
\label{fig-sens}
\end{figure}

\subsubsection{Sensitivity Analysis}

Finally, we turn to investigate the sensitivity of StockODE, e.g., hidden size and the edge rate in HHCN. Note that we do not change other hyperparameters when varying any one investigated hyperparameters.

\textbf{Hidden Size.} As shown in Fig.~\ref{fig:hidden-size} and Fig.~\ref{fig:hidden-size_a}, the left y-axis corresponds to the collected results of SR and NDCF@5 while the right is related to the results of MRR. They demonstrate that the hidden size significantly affects the model performance, especially for MRR. Nevertheless, the larger size of the hidden layer does not bring us better results, and thus we set the hidden size to 64 to trade off the number of trainable parameters and model performance.

\textbf{Relationship Sensitivity.} To quantify the impact of extracted relationships among stocks, we randomly remove 10\% and 50\% relationships from sector-industry relations and Wiki relations, the results in Fig.~\ref{fig:relation} and Fig.~\ref{fig:relation_a} demonstrate that domain-aware dependencies do affect the model performance. And it also indicates that incorporating richer prior knowledge among the stock could bring higher gains in terms of investment decisions.

\textbf{Depth of HHCN.} We turn to study the impact of graph layers in HHCN. Specifically, we increase or decrease the same number of graph layers in intra-domain knowledge and inter-domain knowledge learning, respectively. As Fig.~\ref{fig:Layersofgraph} and Fig.~\ref{fig:Layersofgraph_a} show, we find that using more graph layers does not bring higher achievements. The most plausible reason is the over-fitting issue. Thus, choosing one layer is enough.

\textbf{Depth of Attention Layers.} We finally investigate the impact of attention layers in the Movement Trend Correlation module. As Fig.~\ref{fig:att} and Fig.~\ref{fig:att_a} show, we find a similar observation as the impact of graph layers in HHCN. Hence, we only use a one-layer attention network for movement correlation learning.

\section{Conclusions and Future Work}
This paper presented a novel framework, i.e., StockODE, for stock selection, a practical but intricate task in the field of investment decision-making. In contrast to most existing stock movement prediction efforts, we primarily seek to provide a valuable stock ranking list for investors, aiming to help them alleviate the investment risks. The proposed StockODE inspired by the recent neural dynamic system provides a new perspective on stock movement learning. In particular, we devised a Movement Trend Correlation module to capture the stock relations regarding the time-varying return ratio aspect. Also, we proposed a hierarchical hypergraph to consider both explicit and implicit dependencies among different domains, which aimed to strengthen the impact of higher-order collaboration in the evolution of stocks. Our experimental results demonstrate that StockODE enables promoting the ranking performance compared to state-of-the-art baselines. In the future, we will consider employing financial news and public reviews to provide more evidence impacts behind the stocks.
\label{sec:concl}

\section*{Acknowledgment}
This work was supported by the National Natural Science Foundation of China under Grant 62102326, the Key Research and Development Project of Sichuan Province under Grant 2022YFG0314,  Guanghua Talent Project, and the Financial Intelligence and Financial Engineering Key Laboratory of Sichuan Province.

%



\ifCLASSOPTIONcaptionsoff
  \newpage
\fi



%
\bibliographystyle{IEEEtran}
\bibliography{0_Stock}





\end{document}